%% file: NEW.tex
\documentclass[a4paper,11pt, english]{article} 
\usepackage{jinstpub} 
\input{pool/NextLatexPackagesBase.tex}

\input{pool/NextDefs.tex}

\begin{document}
\title{The Next White (NEW) detector}
\input{NextAuthorListForNewPaper.tex}

\input{src/abstract.tex}


\arxivnumber{1234.56789} 
\collaboration{\includegraphics[height=9mm]{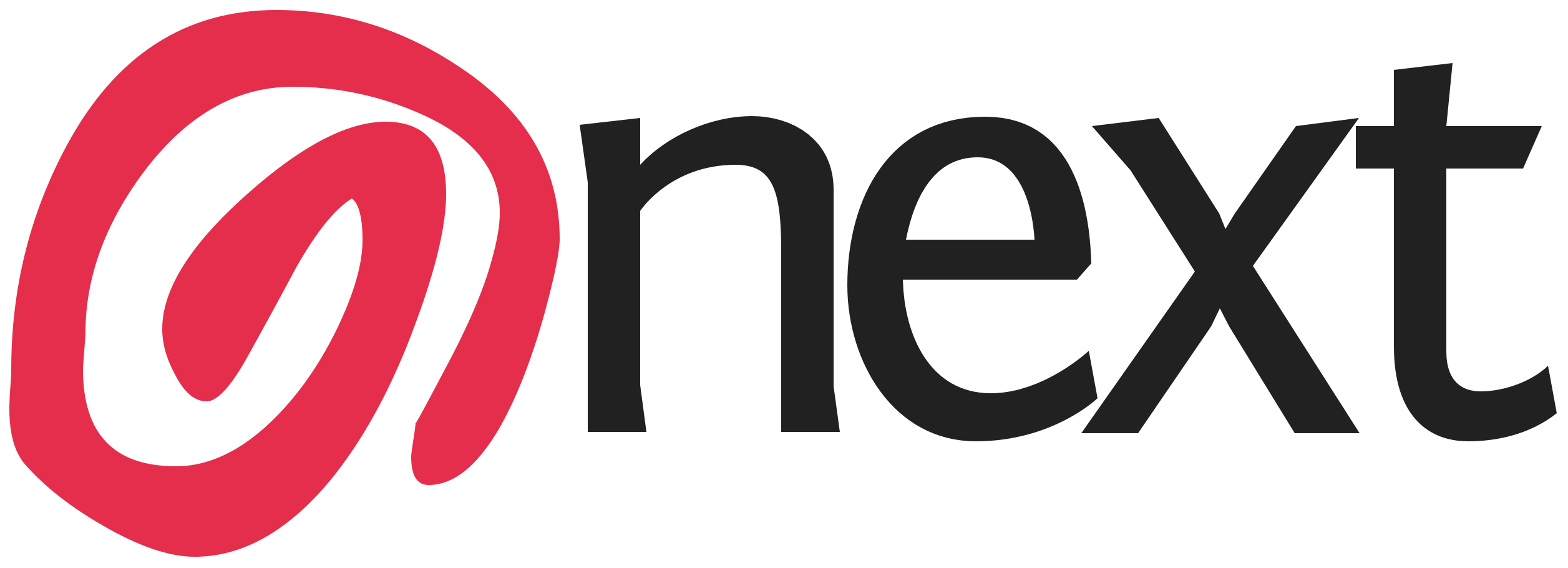}\\[6pt]
The NEXT Collaboration}

\maketitle
\flushbottom

\input{src/introduction.tex}
\input{pool/src/hpxe_principle_of_operation.tex}
\input{pool/src/new_overview.tex}
\input{src/new_tpc.tex}

\input{src/energy_plane.tex}

\input{src/tracking_plane.tex}
\input{src/data_acquistion.tex}

\input{src/gas_system.tex}

\input{src/sensor_calibration.tex}

\input{src/detector_operation.tex}

\acknowledgments
\input{pool/ack2018.tex}

\bibliographystyle{NextRefsStyle}
\bibliography{pool/NextRefs}
\end{document}

%% file: pool/NextLatexPackagesBase.tex
\usepackage{babel}
\usepackage{graphbox}
\usepackage{multirow}
\usepackage{makecell}
\usepackage{amsmath,amssymb, gensymb}
\usepackage{natbib,hyperref}
\usepackage{url}
\usepackage{siunitx}
\usepackage{xspace}
\usepackage{microtype}
\usepackage{subfig}
\usepackage[useregional]{datetime2}

%% file: pool/NextDefs.tex
\newcommand{\NEW}{NEXT-White}
\newcommand{\NEXT}{NEXT-100}

\newcommand{\fig}{figure}

\newcommand{\Fig}{Figure}

\newcommand{\eg}{{\it e.g.}}

\newcommand{\bbonu}{\ensuremath{\beta\beta0\nu}}
\newcommand{\bbtnu}{\ensuremath{\beta\beta2\nu}}

\newcommand{\tz}{\ensuremath{t_0}}
\newcommand{\st}{\ensuremath{S_2}}
\newcommand{\so}{\ensuremath{S_1}}

\newcommand{\NIT}{\ensuremath{N_2}}
\newcommand{\Qbb}{\ensuremath{Q_{\beta\beta}}}

\newcommand{\CS}{\ensuremath{^{137}}Cs}

\newcommand{\NA}{\ensuremath{^{22}}Na}

\newcommand{\KR}{\ensuremath{{}^{82}{\rm Kr}}}

\newcommand{\XE}{\ensuremath{{}^{136}\rm Xe}}

\newcommand{\TL}{\ensuremath{{}^{208}\rm{Tl}}}

\newcommand{\BI}{\ensuremath{{}^{214}}Bi}

\newcommand{\Bi}[1]{\ensuremath{^{#1}\mathrm{Bi}}\xspace}

\DeclareSIUnit\c{\mbox{$c$}}
\DeclareSIUnit\magn{\mbox{$\times$}}
\DeclareSIUnit\min{min}
\DeclareSIUnit\week{week}
\DeclareSIUnit\year{yr}
\DeclareSIUnit\years{years}
\DeclareSIUnit\yr{yr}
\DeclareSIUnit\standard{std}
\DeclareSIUnit\str{sr}
\DeclareSIUnit\ppm{ppm}
\DeclareSIUnit\ppb{ppb}
\DeclareSIUnit\ppt{ppt}
\DeclareSIUnit\pe{PE}
\DeclareSIUnit\spe{SPE}
\DeclareSIUnit\ev{events}
\DeclareSIUnit\ct{counts}
\DeclareSIUnit\neutron{\mbox{$n$}}
\DeclareSIUnit\smp{samples}
\DeclareSIUnit\Sample{S}
\DeclareSIUnit\ch{ch}
\DeclareSIUnit\hit{hit}
\DeclareSIUnit\hits{hits}
\DeclareSIUnit\bin{(\mbox{5-PE}~bin)}
\DeclareSIUnit\sgm{\mbox{$\sigma$}}
\DeclareSIUnit\rms{RMS}
\DeclareSIUnit\keVr{\mbox{keV$_{\rm nr}$}}
\DeclareSIUnit\keVee{\mbox{keV$_{e{\rm e}}$}}
\DeclareSIUnit\ph{photon}
\DeclareSIUnit\pes{pes}
\DeclareSIUnit\el{electrons}
\DeclareSIUnit\pm{PMT}
\DeclareSIUnit\inch{"}
\DeclareSIUnit\bit{bit}
\DeclareSIUnit\sample{samples}
\DeclareSIUnit\barn{barn}
\DeclareSIUnit\bara{bar}
\DeclareSIUnit\barg{barg}
\DeclareSIUnit\mlardepth{\mbox(meter~of~\LAr~depth)}
\DeclareSIUnit\Curie{Ci}
\DeclareSIUnit\psi{psi}
\DeclareSIUnit\parsec{pc}
\DeclareSIUnit\liveday{\mbox{live-days}}
\DeclareSIUnit\days{\mbox{days}}
\DeclareSIUnit\day{\mbox{day}}
\DeclareSIUnit\miles{\mbox{miles}}
\DeclareSIUnit\degreeC{\mbox{$^{\circ}$C}}
\DeclareSIUnit\electron{\mbox{$e^-$}}
\DeclareSIUnit\Euro{\mbox{\euro}}
\DeclareSIUnit\cph{cph}
\DeclareSIUnit\neq{neq}
\DeclareSIUnit\unit{unit}
\DeclareSIUnit\byte{Byte}
\DeclareSIUnit\Bq{\becquerel}

\newcommand{\TPBWaveLength}{\SI{420}{\nano\meter}}

\newcommand{\XeWaveLength}{\SI{172}{\nano\meter}}
\newcommand{\TPBReflectivityInBlue}{\SI{98}{\percent}}

\newcommand{\HPXe}{HPXe}
\newcommand{\HPXeEL}{HPXe-EL}

\newcommand{\TPB}{Tetraphenyl Butadiene}
\newcommand{\PDOT}{Poly Ethylenedioxythiophene}
\newcommand{\ITO}{Indium tin-oxyde}
\newcommand{\RII}{Run II}

\newcommand{\Z}{\ensuremath{z}}

\newcommand{\XY}{\ensuremath{(x, y)}}

\newcommand{\LINNormalTemperature}{\SI{77}{\kelvin}}

\newcommand{\NewGateVoltageSevenBarRunII}{\SI{7.0}{\kV}}
\newcommand{\NewGateVoltageNineBarRunII}{\SI{8.5}{\kV}}
\newcommand{\NewCathodeVoltageRunII}{\SI{27}{\kV}}
\newcommand{\NewCathodeVoltageSevenBarRunII}{\SI{28}{\kV}}
\newcommand{\NewCathodeVoltageNineBarRunII}{\SI{30}{\kV}}

\newcommand{\NewGateVoltageRunII}{\SI{7.2}{\kV}}

\newcommand{\NewCathodeVoltageAtFifteenBar}{\SI{41}{\kV}}

\newcommand{\NewSevenBarPressureRunII}{\SI{7.2}{\bar}}
\newcommand{\NewNineBarPressureRunII}{\SI{9.1}{\bar}}

\newcommand{\NewPressureVesselMaterial}{316Ti}

\newcommand{\NewTpcLength}{\SI{664.5}{\mm}}
\newcommand{\NewTpcBuffer}{\SI{129.5}{\mm}}
\newcommand{\NewTpcDriftLength}{\SI{530.3 +- 2}{\mm}}
\newcommand{\NewResistorsType}{Ohmite HVF 2512}
\newcommand{\NewResistorsValue}{\SI{1}{\giga\ohm}}
\newcommand{\NewResistorsVoltage}{\SI{3}{\kilo\volt}}
\newcommand{\NewFieldCageHDPEThickness}{\SI{21}{\mm}}
\newcommand{\NewCopperRingsPitch}{\SI{1.8}{\mm}}

\newcommand{\NewDriftField}{\SI{400}{\V\per\cm}}

\newcommand{\NewTpcELGap}{\SI{6}{\mm}}
\newcommand{\NewAnodePlateDiameter}{\SI{522}{\mm}}
\newcommand{\NewAnodePlateThickness}{\SI{3}{\mm}}
\newcommand{\NewReducedField}{\SI{2.2}{\kV\per\cm\per\bar}}
\newcommand{\NewReducedFieldRunII}{\SI{1.7}{\kV\per\cm\per\bar}}
\newcommand{\NewELFieldAtFifteenBar}{\SI{27}{\kV\per\cm}}

\newcommand{\NewGateVoltageAtFifteenBar}{\SI{16.2}{\kV}}

\newcommand{\NewNumberOfSiPM}{\num{1792}}
\newcommand{\NewDeadSiPM}{\num{3}}
\newcommand{\NewUnstableSiPM}{\num{6}}
\newcommand{\NewOutSiPM}{\num{4}}
\newcommand{\NewTrackingPlaneActiveAndStable}{99 \%}
\newcommand{\NewNumberOfBoards}{\num{28}}
\newcommand{\NewNumberOfSiPMPerBoard}{8 x 8}
\newcommand{\NewSiPMSeries}{SensL C}
\newcommand{\NewSiPMModel}{MicroFC-10035-SMT-GP}
\newcommand{\NewSiPMSize}{\SI{1}{\mm\square}}
\newcommand{\NewSipmPitch}{\SI{10}{\mm}}
\newcommand{\NewSipmCell}{\SI{35}{\micro\meter}}
\newcommand{\NewSipmDarkCount}{\SI{100}{\kHz}}
\newcommand{\NewPhotoelectronsPerSiPM}{\SI{250}{\pes\per\micro\second}}
\newcommand{\TrackingPlaneToEL}{\SI{8}{\mm}}
\newcommand{\TrackingPlaneToAnode}{\SI{2}{\mm}}
\newcommand{\NewTrackingPlaneEndCapThickness}{\SI{120}{\mm}}
\newcommand{\NewTrackingPlaneConnections}{\num{3600}}
\newcommand{\NewSiPMDataFlow}{\SI{35}{\mega\byte\per\second}}

\newcommand{\NewNumberOfPMT}{\num{12}}
\newcommand{\NewNumberOfCentralPMT}{\num{3}}
\newcommand{\NewCathodeToPMTs}{\SI{130}{\mm}}
\newcommand{\NewPMTDigiSpeed}{\SI{40}{\mega\hertz}}
\newcommand{\NewPMTDataFlow}{\SI{10}{\mega\byte\per\second}}
\newcommand{\NewPMTADCBits}{\num{12}}

\newcommand{\NewTriggerRateBase}{\SI{10}{\hertz}}
\newcommand{\NewMaxTriggerBuffer}{\SI{3.2}{\milli\second}}

\newcommand{\NewTpcDiameter}{\SI{454}{\mm}}
\newcommand{\NewFiducialMass}{\SI{5}{\kg}}
\newcommand{\NewFiducialMassSevenBar}{\SI{2.3}{\kg}}
\newcommand{\NewFiducialMassNineBar}{\SI{3}{\kg}}
\newcommand{\NewPressure}{\SI{15}{\bar}}

\newcommand{\NewPSVolume}{\SI{225}{\liter}}
\newcommand{\NewMainPSVolume}{\SI{140}{\liter}}
\newcommand{\NewGasLoopVolume}{\SI{45}{\liter}}

\newcommand{\NewFirstBottleVolume}{\SI{69}{\liter}}
\newcommand{\NewSecondBottleVolume}{\SI{61}{\liter}}

\newcommand{\NewSecondBottleXenonMass}{\SI{100}{\kg}}
\newcommand{\NewMaxPressure}{\SI{20}{\bar}}
\newcommand{\NewGetterPressure}{\SI{10}{\bar}}
\newcommand{\NewPressureVesselLength}{\SI{950}{\mm}}
\newcommand{\NewPressureVesselDiameter}{\SI{640}{\mm}}

\newcommand{\NewPressureVesselBarrelThickness}{\SI{12}{\mm}}
\newcommand{\NewColdGetters}{SAES MC4500-902}
\newcommand{\NewHotGetters}{SAES PS4-MT50-R-535}
\newcommand{\NewCompressorMinimumInletPressure}{\SI{5}{\bar}}
\newcommand{\NewCompressorMaximumInletPressure}{\SI{10}{\bar}}
\newcommand{\NewCompressorMaximumOutletPressure}{\SI{25}{\bar}}
\newcommand{\CompressorLeakRatePerYear}{\SI{0.19}{\g\per\year}}
\newcommand{\NewExpansionTankVolume}{\SI{2.3}{\cubic\meter}}

\newcommand{\NewPmtEndCapThickness}{\SI{120}{\mm}}

\newcommand{\NewTypePMT}{Hamamatsu R11410-10}
\newcommand{\NewPMTCoverage}{31\%}
\newcommand{\NewOneDotFiveMuFCapacitorsActivity}{\SI{72}{\micro\becquerel\per\unit}}
\newcommand{\NewFourDotSeveMuFCapacitorsActivity}{\SI{123}{\micro\becquerel\per\unit}}
\newcommand{\NewFinechemResistorsActivity}{\SI{4.1}{\micro\becquerel\per\unit}}
\newcommand{\NewBasePinActivity}{\SI{1.1}{\micro\becquerel\per\unit}}
\newcommand{\NewBaseEpoxy}{\SI{1.4}{\milli\becquerel\per\kilogram}}
\newcommand{\NewBaseCable}{\SI{46.8}{\milli\becquerel\per\kilogram}}
\newcommand{\NewBaseSubstrate}{\SI{23}{\micro\becquerel\per\unit}}
\newcommand{\NewBaseCopperCup}{\SI{12}{\micro\becquerel\per\unit}}
\newcommand{\NewKOAResistorsActivity}{\SI{7.7}{\micro\becquerel\per\unit}}
\newcommand{\NewPMTActivity}{\SI{0.35}{\milli\becquerel\per\unit}}
\newcommand{\NewPMTBaseActivity}{\SI{1.2}{\milli\becquerel\per\unit}}

\newcommand{\NewPMTNIT}{\SI{1}{\bar}}

\newcommand{\NewPMTOperatingVoltage}{\SI{1.23}{\kV}}
\newcommand{\NewPMTPhotoelectronEfficiency}{1\%}

\newcommand{\NewPMTGlue}{NyoGel OCK-451}
\newcommand{\NewEnergyPlaneLedPulse}{\SI{50}{\micro\second}}

\newcommand{\NewBarrelICS}{\SI{60}{\mm}}

\newcommand{\NewPlatesICS}{\SI{120}{\mm}}

\newcommand{\NextTpcDiameter}{\SI{1050}{\mm}}
\newcommand{\NextTpcLength}{\SI{1300}{\mm}}


%% file: NextAuthorListForNewPaper.tex
\collaboration{The NEXT Collaboration}
\author[p, j, 1]{F.~Monrabal,} 

\author[j, t, b, 2]{J.J.~G\'omez-Cadenas,}

\author[h]{J.F.~Toledo,}

\author[b]{V.~\'Alvarez,}

\author[b]{J.M.~Benlloch-Rodr\'{i}guez,}

\author[b]{S.~C\'arcel,}

\author[b]{J.V.~Carri\'on,}

\author[h]{R.~Esteve,}
\author[b]{R.~Felkai,}

\author[h]{V.~Herrero,}

\author[b]{A.~Laing,}

\author[b]{A.~Mart\'inez,}

\author[b]{M.~Musti,}

\author[b]{M.~Querol,}

\author[b]{J.~Rodr\'iguez,}

\author[b]{A.~Sim\'on,}

\author[s,2]{C.~Sofka,}

\author[b]{J.~Torrent,}

\author[s]{R.~Webb,}

\author[s,3]{J.T.~White,}

\author[a]{C.~Adams,}

\author[c]{L.~Arazi,}
\author[d]{C.D.R.~Azevedo,}
\author[n]{K. Bailey,}
\author[h]{F.~Ballester,}
\author[e]{F.I.G.M.~Borges,}
\author[b]{A.~Botas,}

\author[f]{S.~Cebri\'an,}
\author[e]{C.A.N.~Conde,}
\author[b]{J.~D\'iaz,}
\author[g]{M.~Diesburg,}
\author[e]{J.~Escada,}
\author[i]{L.M.P.~Fernandes,}
\author[j, t, b]{P.~Ferrario,}
\author[d]{A.L.~Ferreira,}
\author[i]{E.D.C.~Freitas,}
\author[k]{A.~Goldschmidt,}
\author[l]{D.~Gonz\'alez-D\'iaz,}
\author[a]{R.~Guenette,}
\author[m]{R.M.~Guti\'errez,}
\author[n]{K.~Hafidi,}
\author[o]{J.~Hauptman,}
\author[i]{C.A.O.~Henriques,}
\author[m]{A.I.~Hernandez,}
\author[l]{J.A.~Hernando~Morata,}
\author[n]{S.~Johnston,}
\author[p]{B.J.P.~Jones,}
\author[q]{L.~Labarga,}
\author[g]{P.~Lebrun,}

\author[b]{N.~L\'opez-March,}
\author[m]{M.~Losada,}
\author[a]{J.~Mart\'in-Albo,}
\author[l]{G.~Mart\'inez-Lema,}
\author[p]{A.D.~McDonald,}
\author[i]{C.M.B.~Monteiro,}
\author[h]{F.J.~Mora,}
%
%
\author[b]{J.~Mu\~noz Vidal,}
\author[b]{M.~Nebot-Guinot,}
\author[b]{P.~Novella,}
\author[p,1]{D.R.~Nygren,} 
\author[b]{B.~Palmeiro,}
\author[g]{A.~Para,}
\author[b]{J.~P\'{e}rez,}
\author[b]{J.~Renner,}
\author[n]{J.~Repond,}
\author[n]{S.~Riordan,}
\author[b]{C.~Romo,}
\author[r]{L.~Ripoll,}
\author[p]{L.~Rogers,}
\author[e]{F.P.~Santos,}
\author[i]{J.M.F.~dos~Santos,}
\author[b]{M.~Sorel,}
\author[s]{T.~Stiegler,}
\author[d]{J.F.C.A.~Veloso,}
\author[b]{N.~Yahlali}

\note{corresponding author.}
\note{NEXT Co-spokesperson.}
\note{Deceased.}

\emailAdd{monrabal18@gmail.com}

\affiliation[a]{
  Department of Physics, Harvard University\\
  Cambridge, MA 02138, USA}
\affiliation[b]{
Instituto de F\'isica Corpuscular (IFIC), CSIC \& Universitat de Val\`encia\\
Calle Catedr\'atico Jos\'e Beltr\'an, 2, 46980 Paterna, Valencia, Spain}
\affiliation[c]{
  Nuclear Engineering Unit, Faculty of Engineering Sciences, Ben-Gurion University of the Negev\\
P.O.B. 653 Beer-Sheva 8410501, Israel}
\affiliation[d]{
Institute of Nanostructures, Nanomodelling and Nanofabrication (i3N), Universidade de Aveiro\\
Campus de Santiago, 3810-193 Aveiro, Portugal}
\affiliation[e]{
LIP, Department of Physics, University of Coimbra\\
P-3004 516 Coimbra, Portugal}
\affiliation[f]{
Laboratorio de F\'isica Nuclear y Astropart\'iculas, Universidad de Zaragoza\\ 
Calle Pedro Cerbuna, 12, 50009 Zaragoza, Spain}
\affiliation[g]{
Fermi National Accelerator Laboratory\\ 
Batavia, Illinois 60510, USA}
\affiliation[h]{
Instituto de Instrumentaci\'on para Imagen Molecular (I3M), Centro Mixto CSIC - Universitat Polit\`ecnica de Val\`encia\\
Camino de Vera s/n, 46022 Valencia, Spain}
\affiliation[i]{
LIBPhys, Physics Department, University of Coimbra\\
Rua Larga, 3004-516 Coimbra, Portugal}
\affiliation[j]{
Donostia International Physics Center (DIPC)\\
Paseo Manuel Lardizabal 4, 20018 Donostia-San Sebastian.}
\affiliation[k]{
Lawrence Berkeley National Laboratory (LBNL)\\
1 Cyclotron Road, Berkeley, California 94720, USA}
\affiliation[l]{
Instituto Gallego de F\'isica de Altas Energ\'ias, Univ.\ de Santiago de Compostela\\
Campus sur, R\'ua Xos\'e Mar\'ia Su\'arez N\'u\~nez, s/n, 15782 Santiago de Compostela, Spain}
\affiliation[m]
{Centro de Investigaci\'on en Ciencias B\'asicas y Aplicadas, Universidad Antonio Nari\~no\\ 
Sede Circunvalar, Carretera 3 Este No.\ 47 A-15, Bogot\'a, Colombia}
\affiliation[n]
{Argonne National Laboratory,\\ 
Argonne IL 60439, USA}
\affiliation[o]{
Department of Physics and Astronomy, Iowa State University\\
12 Physics Hall, Ames, Iowa 50011-3160, USA}
\affiliation[p]{
Department of Physics, University of Texas at Arlington\\
Arlington, Texas 76019, USA}
\affiliation[q]{
Departamento de F\'isica Te\'orica, Universidad Aut\'onoma de Madrid\\
Campus de Cantoblanco, 28049 Madrid, Spain}
\affiliation[r]{
Escola Polit\`ecnica Superior, Universitat de Girona\\
Av.~Montilivi, s/n, 17071 Girona, Spain}
\affiliation[s]{
Department of Physics and Astronomy, Texas A\&M University\\
College Station, Texas 77843-4242, USA}
\affiliation[t]{
IKERBASQUE, Basque Foundation for Science, 48013 Bilbao, Spain.}

%% file: src/abstract.tex
\abstract{Conceived to host \NewFiducialMass\ of xenon at a pressure of \NewPressure\ in the fiducial volume, the \NEW\ (NEW) apparatus is currently the largest high pressure xenon gas TPC using electroluminescent amplification in the world. It is  also a 1:2 scale model of the \NEXT\ detector scheduled to start searching for \bbonu\ decays in \XE\ in 2019. Both detectors measure the energy of the event using a plane of photomultipliers located behind a transparent cathode. They can also reconstruct the trajectories of charged tracks in the dense gas of the TPC with the help of a plane of silicon photomultipliers located behind the anode. A sophisticated gas system, common to both detectors, allows the high gas purity needed to guarantee a long electron lifetime. \NEW\ has been operating since October 2017 at the Canfranc Underground Laboratory (LSC), in Spain. This paper describes the detector and associated infrastructures.}

%% file: src/introduction.tex
\section{Introduction}
\label{sec:intro}

\input{pool/src/intro_new.tex}

This paper is organized as follows: 
section \ref{sec.hpxe} discusses the principle of operation of a high-pressure xenon TPC with electroluminescent readout (\HPXeEL); section \ref{sec.new} offers an overview of the detector and the associated infrastructures at the LSC; sections \ref{sec.tpc}, \ref{sec.ep} and \ref{sec.tp} detail the three main subsystems: time projection chamber, energy plane and tracking plane, while \ref{sec.daq} describes the data acquisition and trigger. The gas system developed  for NEW (and NEXT-100), and whose early commissioning was one of the main reasons to build NEW is summarized in section \ref{sec.gas}. Following the detector's description, section \ref{sec.calib} explains sensor calibration. To conclude, section \ref{sec.operation} summarizes the main instrumental lessons learned during the first year of detector operation.
 

%% file: pool/src/intro_new.tex
The NEXT program is developing the technology of high-pressure xenon gas Time Projection Chambers (TPCs) with electroluminescent amplification (\HPXeEL) for neutrinoless double beta decay searches \cite{Nygren:2009zz, Alvarez:2011my, Alvarez:2012haa, Gomez-Cadenas:2013lta, Martin-Albo:2015rhw}. 
The first phase of the program included the construction, commissioning and operation of two prototypes, called NEXT-DEMO and 
NEXT-DBDM, which demonstrated the robustness of the technology, its excellent energy resolution and its unique topological signal \cite{Alvarez:2012xda, Alvarez:2013gxa, Alvarez:2012hh, Ferrario:2015kta}.
 
The \NEW\footnote{Named after Prof.~James White, our late mentor and friend.} (NEW) detector implements the second phase of the program. \NEW\ is a $\sim$ 1:2 scale model of NEXT-100 (the TPC has a length of \NewTpcLength\ and a diameter of \NewAnodePlateDiameter\, while the NEXT-100 TPC has a length of \NextTpcLength\ and a diameter of  \NextTpcDiameter), a 100 kg \HPXeEL\ detector, which constitutes the third phase of the program and is foreseen to start operations in 2019. \NEW\ has been running successfully since October 2016 at Laboratorio Subterr\'aneo de Canfranc (LSC). Its purpose is to validate the \HPXeEL\ technology in a large-scale radiopure detector.
This validation is composed of three main tasks: to assess the robustness and reliability of the technological solutions;
to compare in detail the background model with data, particularly the contribution to the radioactive budget of the different components, and
to study the energy resolution and the background rejection power of the topological signature characteristic of a \HPXeEL. 
Furthermore, \NEW\ can provide a measurement of the two-neutrino double beta decay mode (\bbtnu). 
\par

%% file: pool/src/hpxe_principle_of_operation.tex
\section{Principle of operation of \HPXeEL\ TPCs}
\label{sec.hpxe}

\begin{figure}[bhtp!]
\centering
\includegraphics[width=0.8\textwidth]{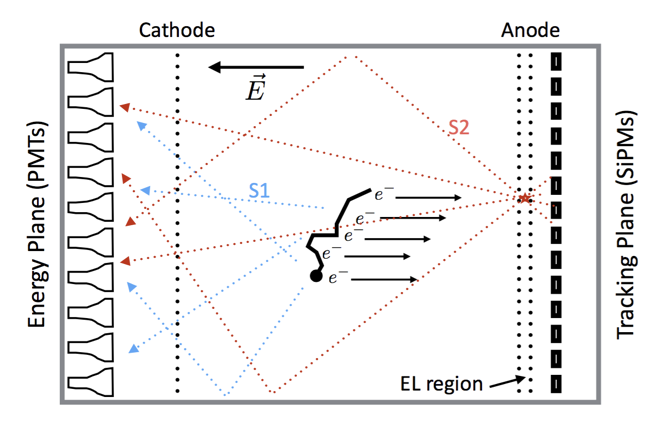}
\caption{\small Principle of operation of an \HPXeEL.} 
\label{fig.po}
\end{figure}

\Fig\ \ref{fig.po} illustrates the principle of operation of a high pressure xenon gase \HPXe. Charged tracks propagating in the TPC
lose energy  by ionizing and exciting the gas atoms. These, in turn, will self-trap to form excimer states that de-excite by emitting vacuum ultraviolet (VUV) photons of \XeWaveLength\ wavelength. This early scintillation is the so-called primary signal (\so). The secondary electrons produced by ionization along the track path will be drifted, under the influence of an electric field, towards the TPC anode. Near the anode, this delayed signal can be amplified by a variety of mechanisms. In the case of a \HPXeEL, the amplification is performed using electroluminescence (EL) \cite{Aprile:2008bga}). An intense electric field is created in a region between the TPC gate held at high voltage and the grounded anode, accelerating the ionization electrons sufficiently to produce \st\ scintillation light. Yet, the EL field is kept low enough to avoid additional gas ionization, which would harm the linearity of the amplification process.

An  \HPXeEL\ is, therefore, an {\em optical TPC}, where both scintillation and ionization produce a light signal. In the NEXT detectors, the light is detected by two independent sensor planes located behind the anode and the cathode. The energy of the event is measured by integrating the amplified EL signal (\st) with a plane of photomultipliers (PMTs). This {\em energy plane} also records the \so\ signal which provides the start-of-event (\tz).  

Electroluminescent light allows tracking, since it is detected as well a few mm away from production at the anode plane, via a dense array of silicon photomultipliers (SiPMs), which constitute the  \emph{tracking plane}. As \st\ light is produced in the EL region, near the sensors, \XY\ information transverse to the drift direction can be extracted from the SiPM response. The longitudinal (or \Z) position of the event can instead be obtained through the drift time, which is defined as the time difference between the \so\ and \st\ signals. In other words, the drift time is the time elapsed between the ionization time of the gas (marked by \so) and the time when the electrons reach the EL region (marked by \st). 

%% file: pool/src/new_overview.tex
\section {The \NEW\ detector: an overview}
\label{sec.new}

\begin{table}[htp]
\caption{\NEW\ TPC parameters.}
\begin{center}
\begin{tabular}{|c|c|c|c|}
\hline
TPC parameter & Nominal & \RII (4734) & \RII (4841) \\
\hline
Pressure & \NewPressure & \NewSevenBarPressureRunII & \NewNineBarPressureRunII \\
EL field (E/P) & \NewReducedField & \NewReducedFieldRunII & \NewReducedFieldRunII \\
EL gap & \NewTpcELGap & \NewTpcELGap & \NewTpcELGap \\
$V_{gate}$ & \NewGateVoltageAtFifteenBar & \NewGateVoltageSevenBarRunII & \NewGateVoltageNineBarRunII\\
Length & \NewTpcLength & \NewTpcLength & \NewTpcLength \\
Diameter &  \NewTpcDiameter & \NewTpcDiameter & \NewTpcDiameter \\
Fiducial mass & \NewFiducialMass & \NewFiducialMassSevenBar & \NewFiducialMassNineBar\\
Drift length & \NewTpcDriftLength & \NewTpcDriftLength & \NewTpcDriftLength \\
Drift field & \NewDriftField & \NewDriftField& \NewDriftField \\
$V_{cathode}$ &\NewCathodeVoltageAtFifteenBar &  \NewCathodeVoltageSevenBarRunII & \NewCathodeVoltageNineBarRunII\\
\hline\hline
\end{tabular}
\end{center}
\label{tab.TPC}
\end{table}%

The \NEW\ apparatus is currently the world's largest radiopure \HPXeEL. Table \ref{tab.TPC} shows the main parameters of the TPC. The left column lists the nominal (design) parameters, while the center and right columns list the operational parameters during the initial operation of the detector (the so-called \RII\ described later in this paper). The
energy plane is instrumented with \NewNumberOfPMT\ \NewTypePMT\  PMTs located \NewCathodeToPMTs\ behind the cathode, providing a coverage of \NewPMTCoverage. The tracking plane is instrumented with \NewNumberOfSiPM\ SiPMs SensL series-C distributed at a pitch of \NewSipmPitch. An ultra-pure copper shell (ICS) \NewBarrelICS\ thick, acts as a shield in the barrel region. The tracking plane and the energy plane are supported also by pure copper plates \NewPlatesICS\ thick.  

 \begin{figure}[bhtp!]
\centering
\includegraphics[width=0.9\textwidth]{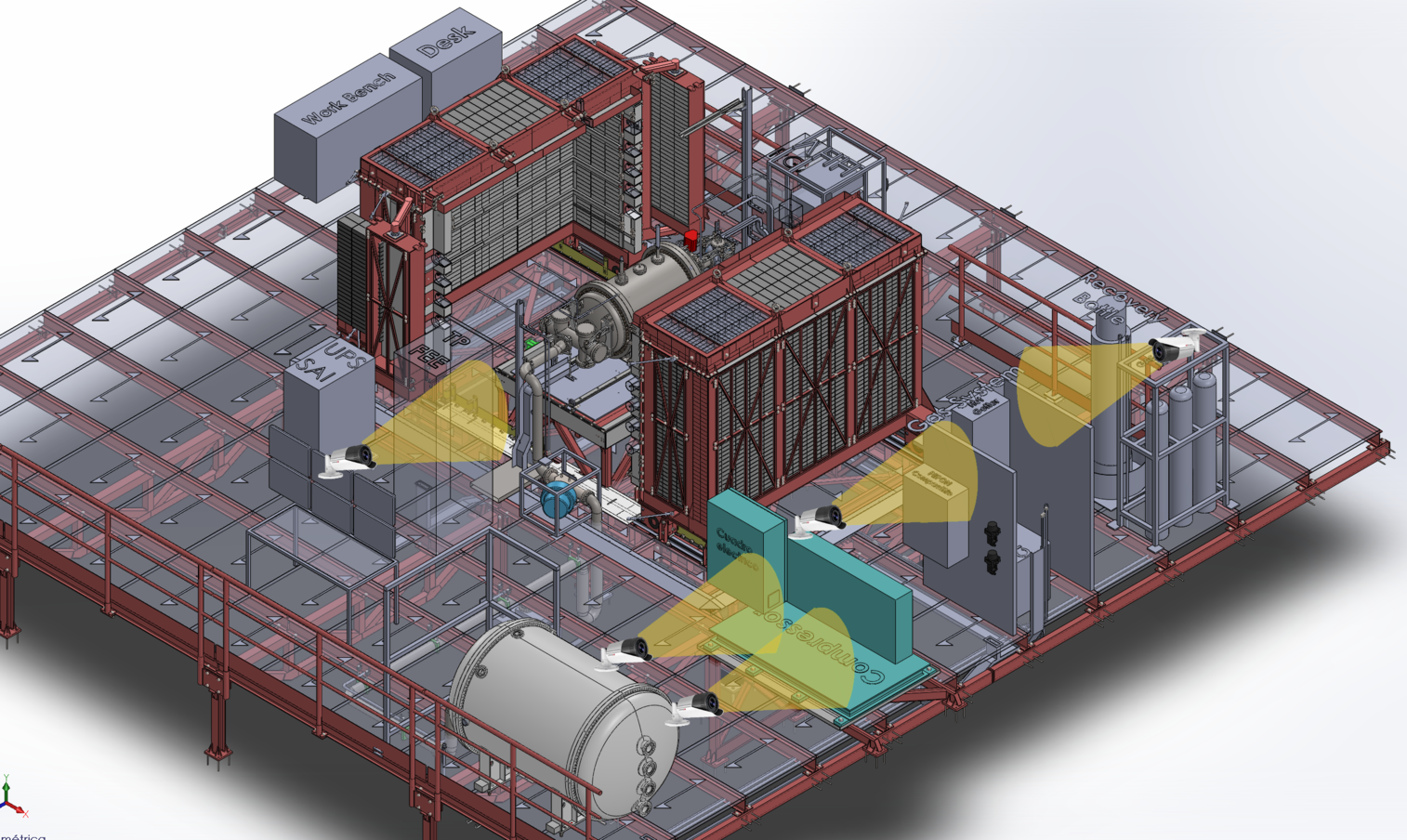}
\includegraphics[width=0.9\textwidth]{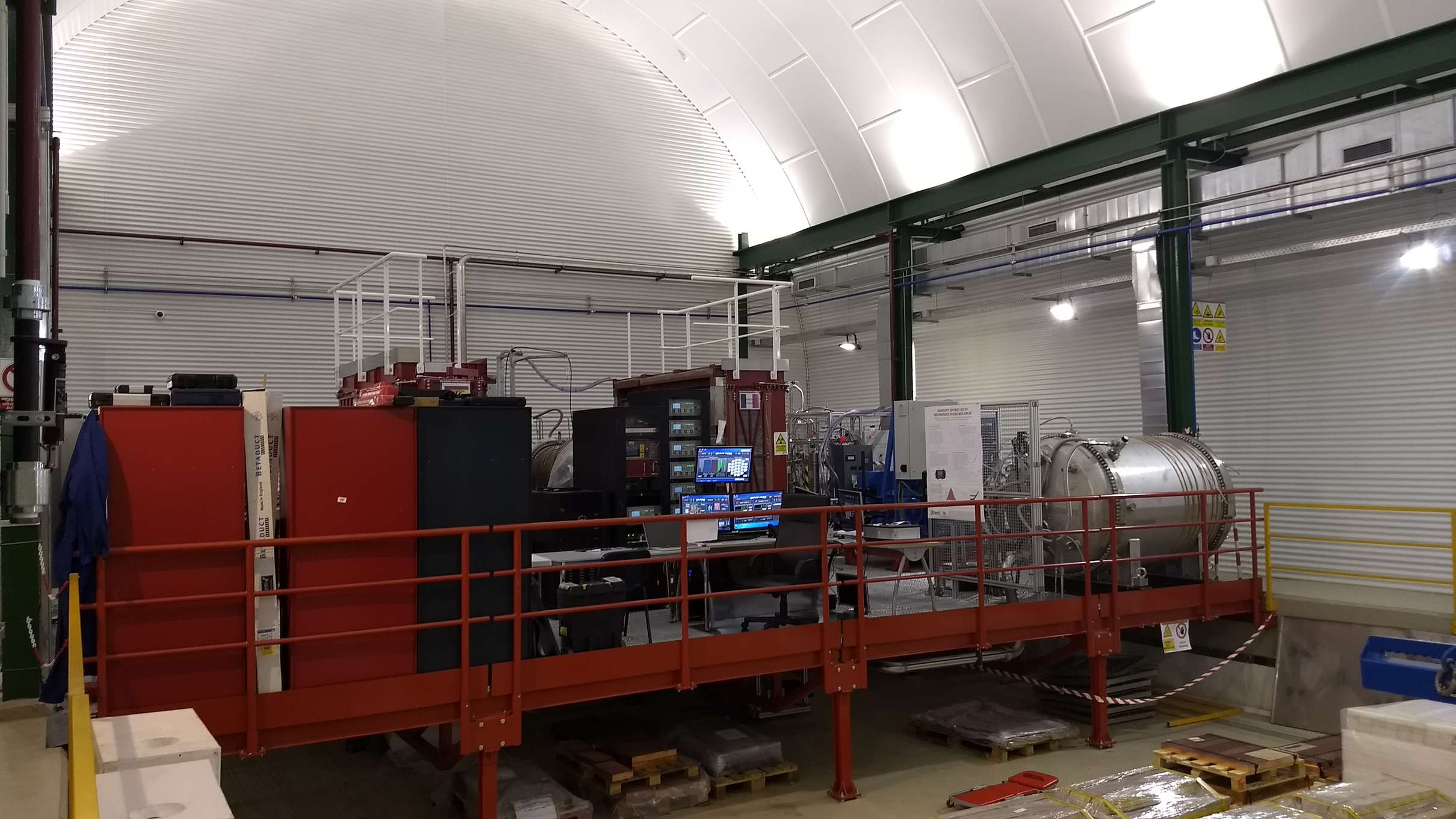}
\caption{\small Top panel: a diagram showing the NEW detector and associated infrastructures at the LSC (the yellow beams mark the coverage of webcams for remote parameter control); bottom panel: a picture of the setup at the LSC.} \label{fig.newatlsc}
\end{figure} 
 
The detector operates inside a pressure vessel fabricated with a radiopure titanium alloy, \NewPressureVesselMaterial. The pressure vessel sits on a seismic table and is surrounded by a lead shield (the lead castle). Since a long electron lifetime is a must, the xenon circulates in a gas system where it is continuously purified. The whole setup sits on top of a tramex platform elevated over the ground at Hall-A, in the Laboratorio Subterr\'aneo de Canfranc (LSC) (figure \ref{fig.newatlsc}). 

%% file: src/new_tpc.tex
\section{The NEW TPC}
\label{sec.tpc}

\begin{figure}[!htb]
\centering
\includegraphics[width=0.8\textwidth]{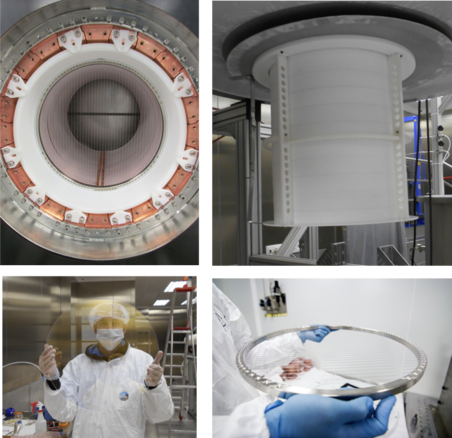}
\caption{\small The main components of the TPC. Top left, the field cage, showing the HDPE body, the copper rings and the cathode grid; bottom left, the anode quartz plate; top right, the light tube; bottom right, the gate grid.} \label{fig.TPC}
\end{figure} 

Structurally, the \NEW\ TPC (figure \ref{fig.TPC}) consists of the field cage (FC), cathode and gate grids, anode plate, and high voltage feedthroughs (HVFT). Functionally, the TPC includes the drift region (DR), buffer region (BR) and the electroluminescent region (ELR).  

\subsection{Field Cage and ELR}

\begin{figure}[!htb]
\centering
\includegraphics[angle=0, width=0.65\textwidth]{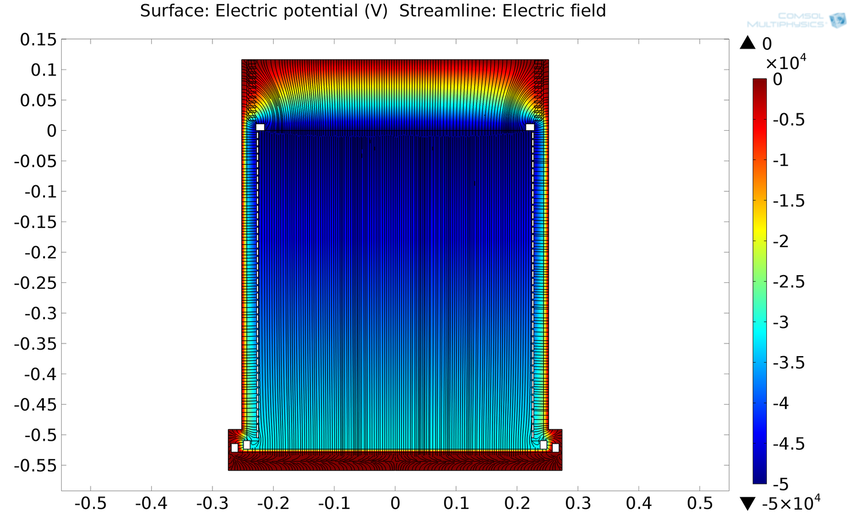}
\includegraphics[angle=0, width=0.45\textwidth]{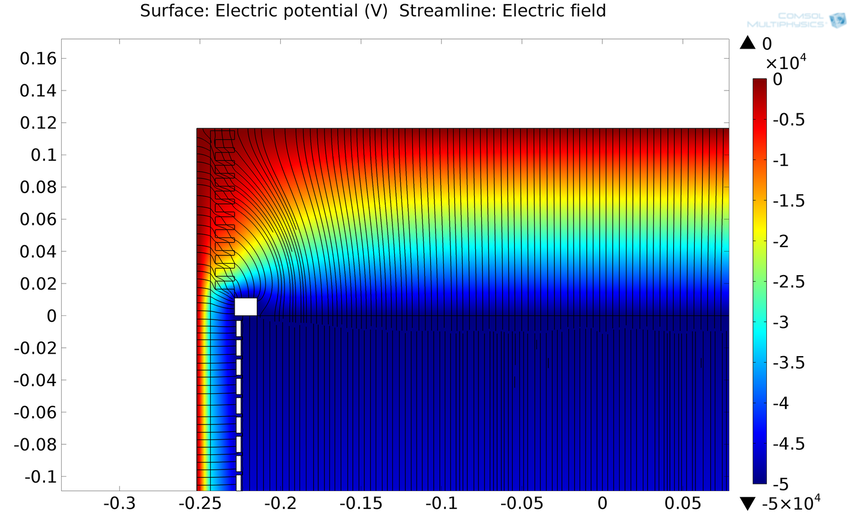}
\includegraphics[angle=0, width=0.45\textwidth]{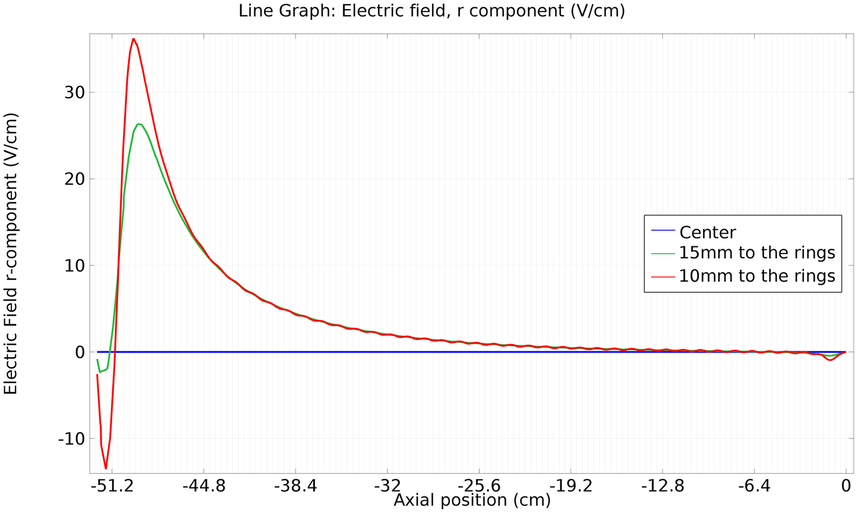}
\caption{\small Simulation of the electric field in the \NEW\ field cage. Top: General view of the electric field and voltage (color). Bottom left: Detail of the electric field on the buffer region, for the general simulation the grooves were not rounded in order to reduce computing time. Bottom right: Radial component of the electric field through different axial lines. } \label{fig:fc_comsol_full}
\end{figure}

The field cage body is a high density Polyethylene (HDPE) cylindrical shell of 
\NewFieldCageHDPEThickness\ thickness. The inner part of the field cage body is machined to produce grooves where radiopure copper rings are inserted
(figure \ref{fig.TPC}, top left panel). 
The drift field is created by applying a voltage difference between the cathode and the gate, which are transparent steel grids mounted on radiopure steel frames (figure \ref{fig.TPC}, bottom right panel). The field transports the ionization electrons to the anode where they are amplified. Good uniformity is necessary to  preserve the topological information, avoid charge losses and prevent charging up of the field cage walls. The role of the copper rings is to correct the field to guarantee high uniformity all across the field cage. The rings have the shape of a rounded rectangle of \SI{10 x 3}{mm} and radius \SI{0.5}{mm} on the edges and are mounted at a pitch of
\NewCopperRingsPitch\ in the inner part of the HDPE support. They are connected through a pair of radio-pure resistors, 
 \NewResistorsType\ of \NewResistorsValue\ each, rated to operate at \NewResistorsVoltage\ with 1\% tolerance. 
 
 The electric field has been simulated using COMSOL\textsuperscript{\textregistered} finite element software (figure \ref{fig:fc_comsol_full} bottom). A 2D-axis symmetric approximation was used (figure \ref{fig:fc_comsol_full}). Radial components of the electric field have been evaluated through different axial positions. The field uniformity is satisfactory for a radial distance up to 20 mm from the walls.
 
The buffer region between the cathode and the energy plane guarantees that the 
large cathode voltage can be smoothly degraded to the ground voltage at which the PMTs' photocathodes sit. The buffer has a length of \NewTpcBuffer, almost 5 times shorter than the active volume, and therefore with an electric field five times larger. On the other hand, the  field in this region does not need to be homogeneous, and therefore there is no need to use field rings. The voltage is then downgraded allowing enough distance between the cathode and the energy plane so that the electric field is well below the breackdown values of the gas and the EL threshold. The HDPE body in the BR was grooved in both the inner and the outer side to impede charge movements.

The amplification or electroluminescent region is the most delicate part of the detector, given that the electric field is very high (\NewELFieldAtFifteenBar at \NewPressure) and at the same time it must be very uniform, since field inhomogeneities deteriorate the energy resolution. The anode is defined by a \ITO\ (ITO) surface coated over a fused silica plate of \NewAnodePlateDiameter\ diameter and \NewAnodePlateThickness\ thickness (figure \ref{fig.TPC}, bottom left panel). The whole region is mounted on top of the tracking plane to ensure its flatness and it is only connected to the rest of the field cage when closing the detector. A thin layer of \TPB\ (TPB), commonly used in noble gases detectors to shift VUV light to the visible spectrum \cite{Gehman:2011xm} is vacuum-deposited on top of the ITO. 

\subsection{High Voltage feedthroughs}
\label{subsec:fc_hvft}

\begin{figure}[!htb]
\centering
\includegraphics[angle=0, width=0.48\textwidth]{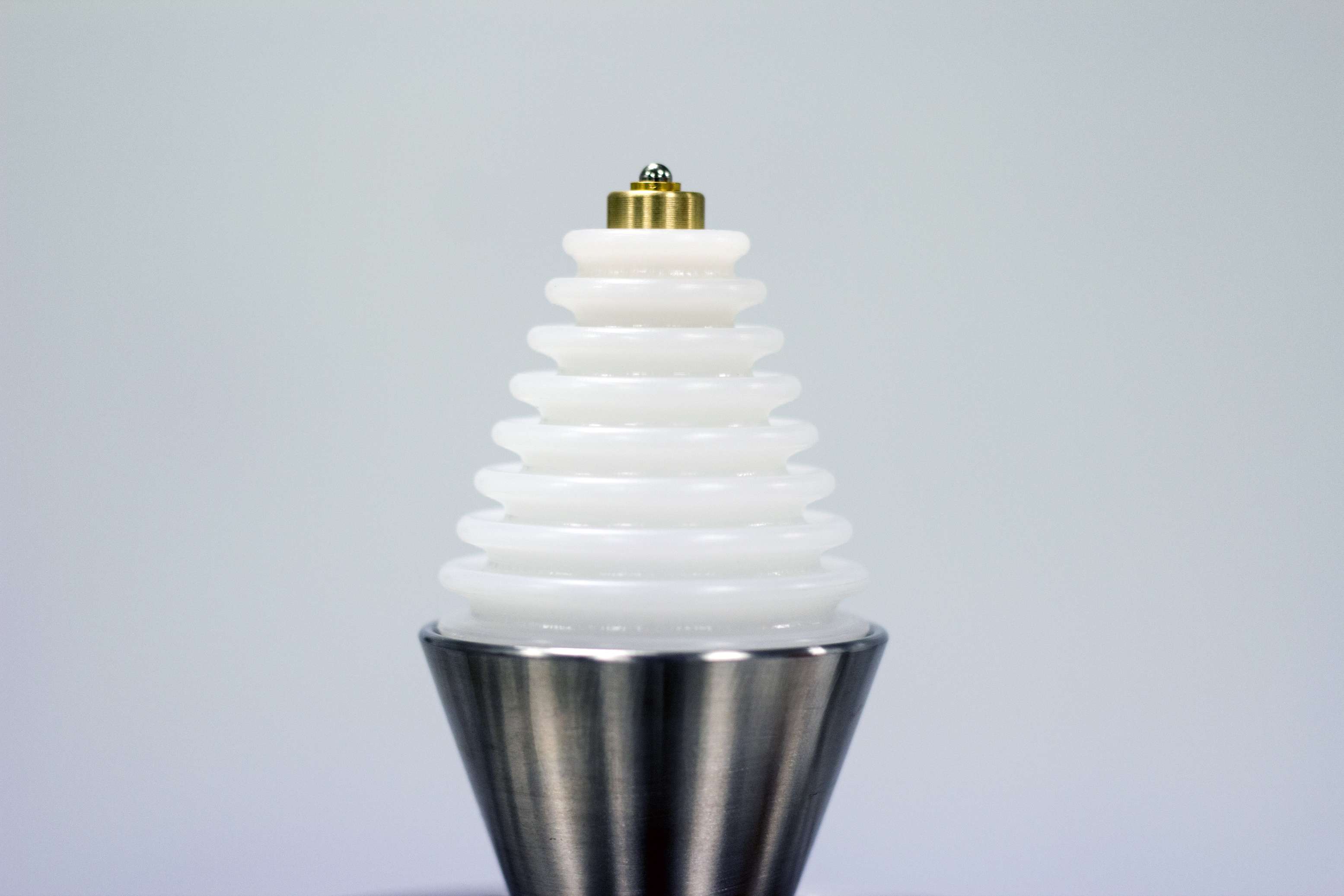}
\includegraphics[angle=0, width=0.42\textwidth]{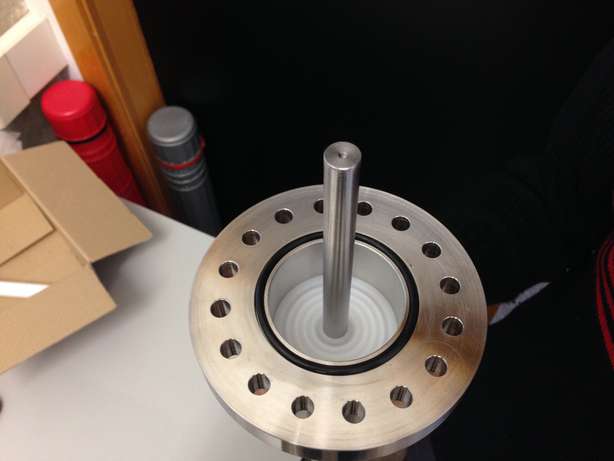}
\caption{\small Left: the conic-design, cathode HVFT. Right: the gate HVFT. } 
\label{fig:fc_hvft}
\end{figure}

The \NEW\ high voltage feedthroughs (HVFT) must hold a high voltage (e.g., the design parameters are \NewCathodeVoltageAtFifteenBar\ at the cathode and 
\NewGateVoltageAtFifteenBar\ at the gate). This is made difficult by the fact that noble gases are poor insulators \cite{CHRISTOPHOROU1988424}, \cite{VIJH1985287} being surface discharges a very important effect. Furthermore, commercial feedthroughs are usually fabricated using non radiopure materials (such as ceramics).  The design and construction of custom-made radiopure HVFT capable to hold high voltages in xenon has been one of the most challenging efforts of the \NEW\ project.

Two different HVFT have been built for the cathode and the gate as shown in 
figure \ref{fig:fc_hvft}.
The cathode feedthrough \cite{Rebel:2014uia}\footnote{based on the ideas of H. Wang, of UCLA} implements a conic design in order to prevent electric field pileup in the region near ground while at the same time forcing the electric field to be parallel to the dielectric surface, thus preventing charge accumulation.
The design of the gate feedthrough is much simpler, with a straight metal rod touching the gate, so that the gas itself provides the needed insulation, avoiding dielectric material near the end of the electrode and thus preventing surface sparks. The dielectric is placed at HVFT entry point in the chamber were the electric field is already parallel to the surface. 
Both feedthroughs are fabricated with HDPE and cryo-fitted in an external piece made of radiopure stainless steel.

\subsection{Light Tube}


Collection of light in an optical TPC is crucial, and thus the TPC needs to be surrounded by a light-tube capable of reflecting light emitted at large angles so that as many photons as possible end up hitting the cathode PMTs. The light tube is made of teflon coated with TPB in order to maximize light collection, since the reflectivity of teflon at the peak emission wavelength of the TPB (\TPBWaveLength) is very 
high (\TPBReflectivityInBlue) \cite{Boccone:2009kk}. The top-right panel of figure \ref{fig.TPC} shows the light tube, made of several teflon rings (each ring has a height of 15 cm), placed in the support of the large TPB evaporator (part of the facilities of the Dark Side experiment, at the Laboratory of Gran Sasso, LNGS). The light tube was coated with TPB at the LNGS before installation in the \NEW\ detector.

%% file: src/energy_plane.tex
\section{Energy plane}
\label{sec.ep}

\begin{figure}[!htbp]
\centering
\includegraphics[angle=0, width=0.48\textwidth]{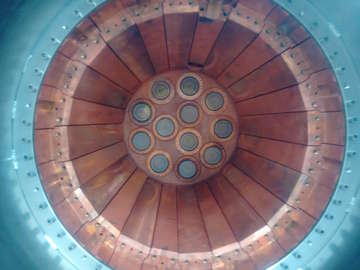}
\includegraphics[angle=0, width=0.48\textwidth]{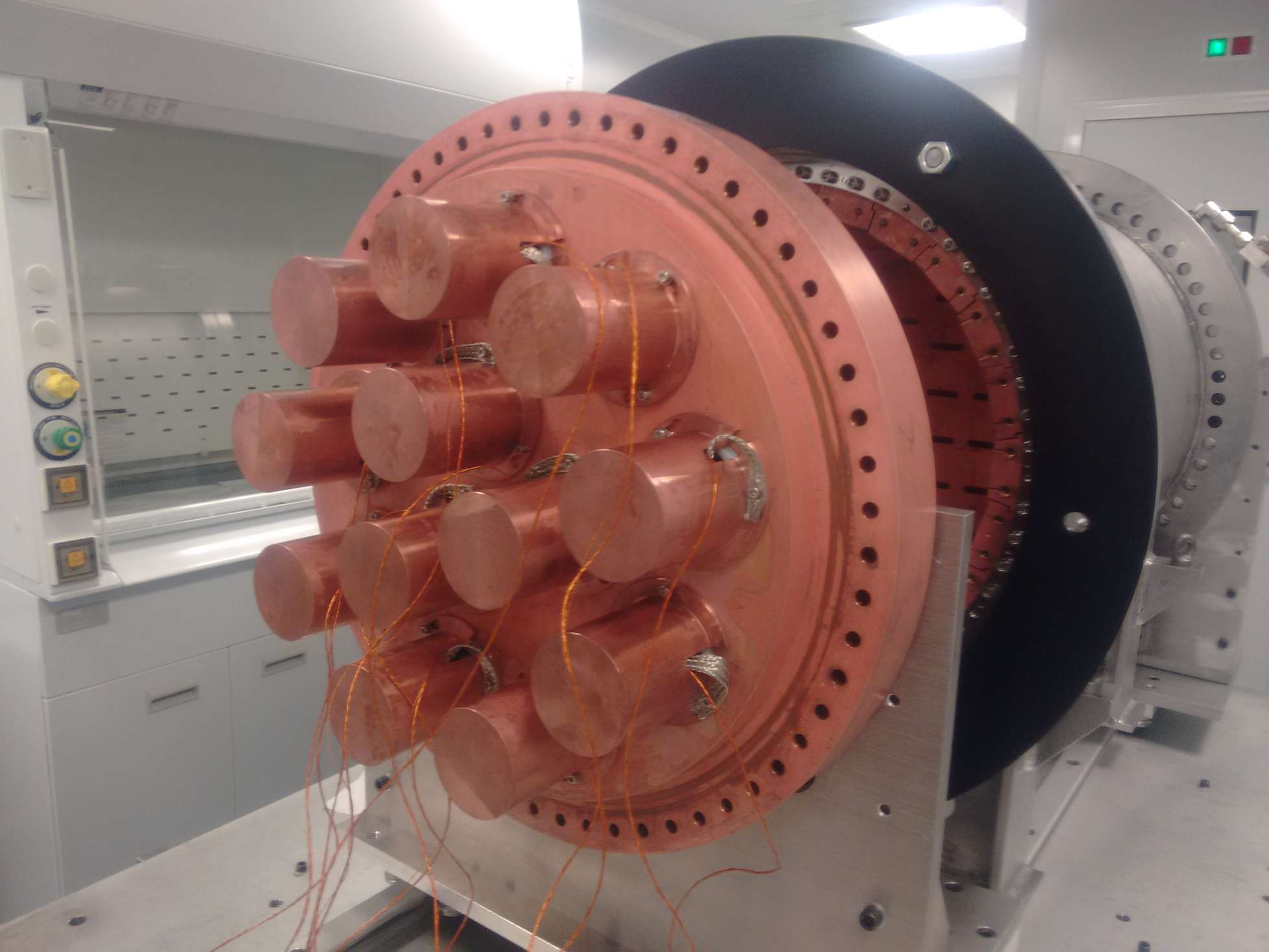}
\includegraphics[angle=0, width=0.48\textwidth]{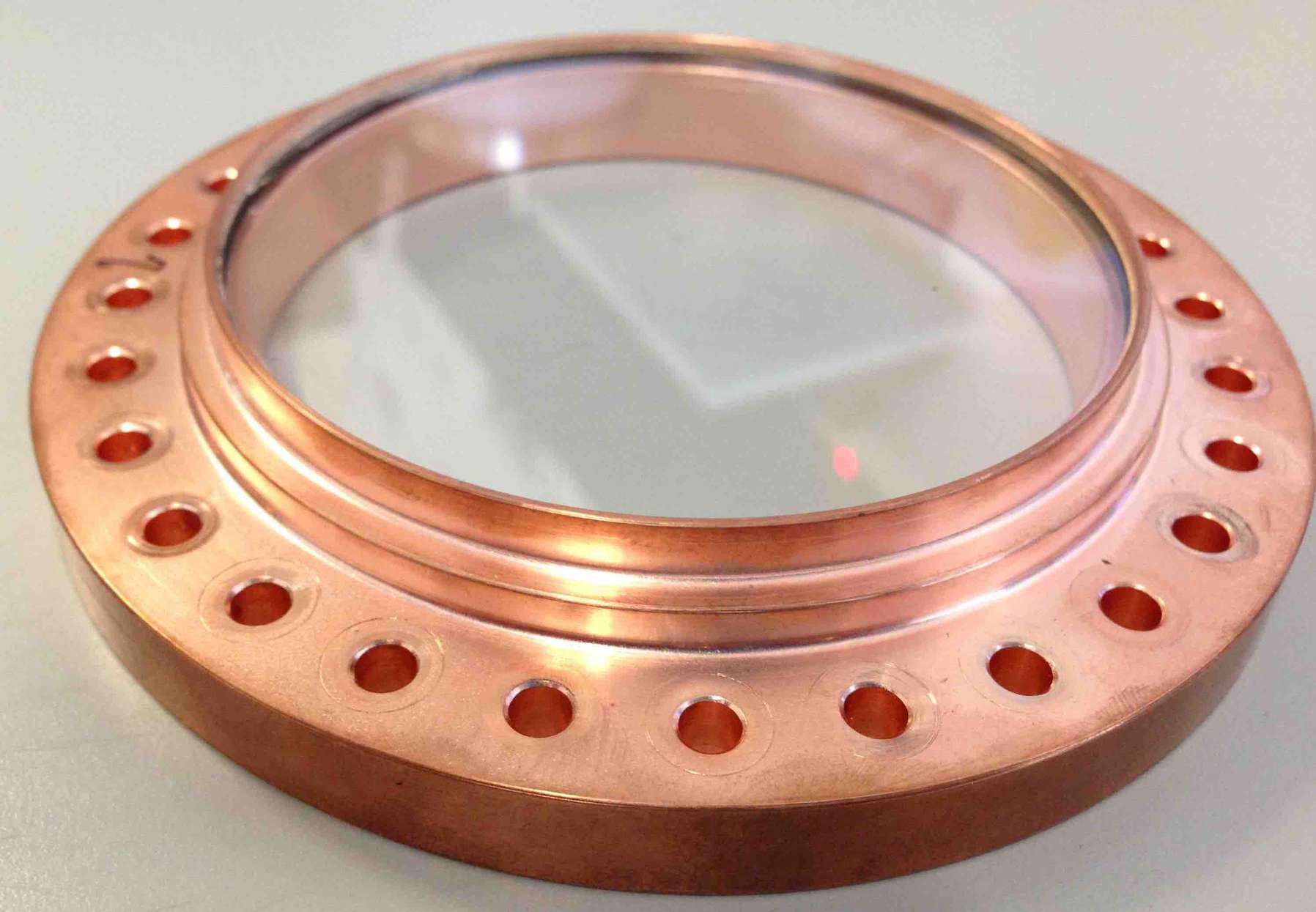}

\caption{\small Left panel: the energy plane viewed from the anode, showing the sapphire windows coated with PEDOT and TPB; right panel: The copper hats that protect the PMTs and shield from external radiation.  Bottom: Detail of the copper frame with the sapphire window.} \label{fig.EP}
\end{figure}

The measurement of the event energy as well as the detection of the primary scintillation signal that determines the \tz\ of the event is performed by the \NEW\ energy plane (EP), shown in figure \ref{fig.EP}. The \NewTypePMT\ PMTs are chosen for their low radioactivity
(\NewPMTActivity) \cite{Cebrian:2017jzb} 
and good performance~ \cite{HamamatsuPMTs}). Since they cannot withstand high pressure they are protected from the main gas volume by a radiopure copper plate \NewPmtEndCapThickness\ thick, which also acts as a shielding against external radiation. 
The PMTs are coupled to the xenon gas volume through \NewNumberOfPMT\ sapphire windows welded to a radiopure copper frame that seals against the copper plate (figure \ref{fig.EP} bottom).
The windows are coated with a resistive (and very transparent) compound: \PDOT\ (PEDOT) \cite{sigma:pedot} in order to set the windows surfaces to ground while at the same time avoiding sharp electric field components near the PMT windows. A thin layer of TPB is vacuum-deposited on top of the PEDOT.

The PMTs are optically coupled to the sapphire window using \NewPMTGlue.
The copper plate end-cap creates an independent volume that allows operation of the PMTs in vacuum or at low pressure.  Currently the PMTs are operated in a \NIT\ atmosphere at 
 \NewPMTNIT.

The \NewNumberOfPMT\ PMTs are distributed in a circular group of external PMTs 
with \NewNumberOfCentralPMT\ in the center. The total photocathode coverage of \NewPMTCoverage, implying that 
\NewPMTPhotoelectronEfficiency\ of the photons produced in the EL region results in a photoelectron. 

\begin{table}[htp!]
\caption{Radioactive \BI\ budget of \NEW\ PMTs and base circuits.}
\begin{center}
\begin{tabular}{|c|c|}
\hline
Component & \Bi\ activity \\
\hline\hline
PMT &  \NewPMTActivity\\
\hline
Base & \\ 
Capacitors 1.5 microF & \NewOneDotFiveMuFCapacitorsActivity \\
Capacitors 4.7 microF & \NewFourDotSeveMuFCapacitorsActivity\\
Finechem resistors & \NewFinechemResistorsActivity\\
KOA RS resistors & \NewKOAResistorsActivity\\
Pin receptacles & \NewBasePinActivity \\
Araldite epoxy & \NewBaseEpoxy\\
Kapton-Cu cable &\NewBaseCable\\
Kapton substrate &\NewBaseSubstrate\\
Copper cap &\NewBaseCopperCup\\
\hline
Total base  & \NewPMTBaseActivity \\ 
\hline
\end{tabular}
\end{center}
\label{tab.PMTBudget}
\end{table}%

\subsection{PMT base circuit}

\begin{figure}[!htbp]
\centering
  \includegraphics[width=\textwidth]{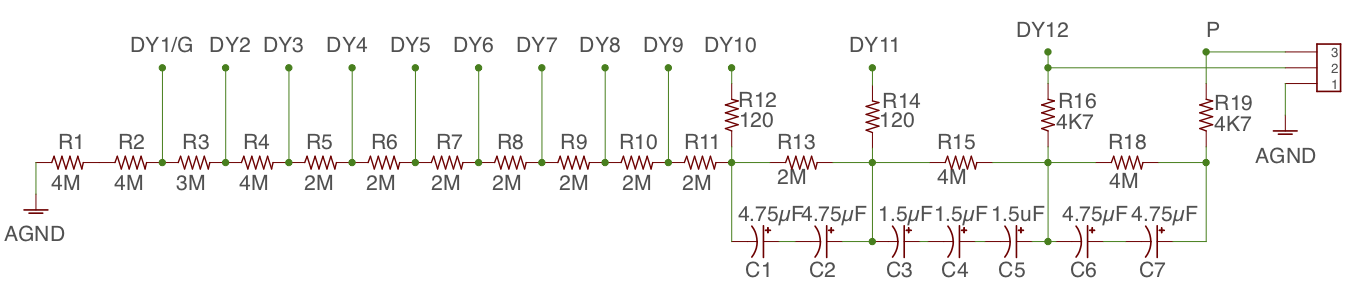}
 
\caption{\small  \NEW\ PMT base circuit.}
\label{fig:7C}
\end{figure}

%
%
The PMT base, shown in \fig\ \ref{fig:7C} is a passive circuit involving 19 resistors of different electrical resistance, 7 capacitors (5 having a capacitance 
of \SI{1.5}{\micro\farad} and 2 with  \SI{4.7}{\micro\farad}) and 18 pin receptacles
soldered on a kapton circuit board and covered with thermal epoxy (Araldite 2011 (R) ) to avoid dielectric breakdown in moderate vacuum or in a N2 atmosphere, this epoxy also guarantees good thermal contact with the copper hat in case of operation at vacuum.
 It has been designed to optimize the tradeoff  between the conflicting targets of preserving the signal linearity and keeping the radioactive budget to a minimum. The difficulty to achieve both goals simultaneously resides in the fact that the stringent requirements to keep a linear response ---the design  demands that the maximum expected signal introduces a voltage drop no larger than 0.1\% of the voltage between dynodes--- requires the use of capacitors to hold the charge in the latest amplification stages, where the gain is very large. The linearity fit gives a 0.38\% maximum deviation for an input of \num{140E+3} photoelectrons which exceeds the maximum charge expected per time bin (25 ns) in the PMTs. 


The tradeoff stems from the radioactivity introduced by the base, which is four times larger than that of the PMT itself, as shown, for \BI\ activity in table \ref{tab.PMTBudget} (see ~\cite{Cebrian:2017jzb} for a thorough discussion). Notice, however, that the radioactivity injected in \NEW\ by the base is partially attenuated by the copper shield, and contributes roughly the same to the final background count as the PMTs. 

The resistor chain value ratios have been recommended by the PMT manufacturer and the final value of the resistors have been adjusted to reduce power dissipation. A power of 40 mW results in a stable temperature of \SI{30}{\degreeC} (the ambient temperature at the LSC is quite stable at \SI{23}{\degreeC}) which introduces no disturbances in the rest of the detector. The number of capacitors ---which are the most radioactive elements in the base-- have been kept to a bare minimum. They are only introduced in the last three amplification stages where the amount of charge to be delivered by the PMT is higher and needs to be held with the help of capacitors. 

\subsection{Front End Electronics}


\begin{figure}[!htb]
\centering
\includegraphics[width=\textwidth]{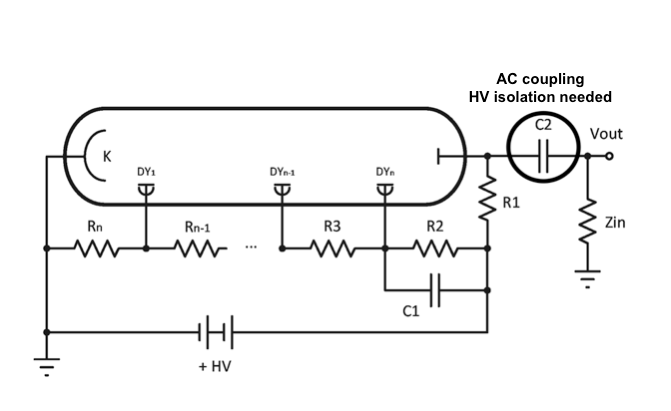}
\caption{\small Grounded cathode PMT connection scheme.}
\label{fig:grounded_cathode}
\end{figure}


In \NEW\ the PMTs' photocathodes (and thus the PMT's bodies) are connected to ground, while  the anode is set at the operating voltage of the PMT (\NewPMTOperatingVoltage).
This solution simplifies the detector mechanics and enhances safety (the alternative, with the cathode and the PMT body at high voltage would have required to isolate each PMT from ground). In exchange, the anode output needs to be AC coupled through a decoupling high-voltage capacitor, as shown in  \fig\ \ref{fig:grounded_cathode}.


%
%
%

\begin{figure}[!htbp]
\centering
\includegraphics[width=0.5\textwidth]{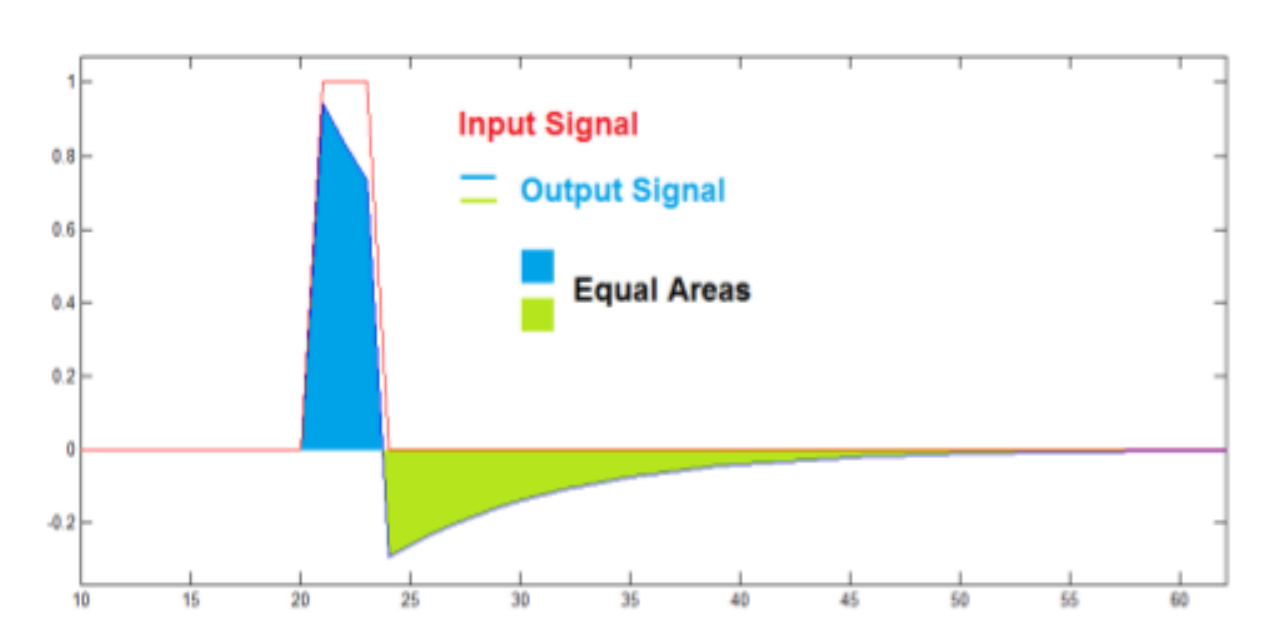} 
\includegraphics[width=0.5\textwidth]{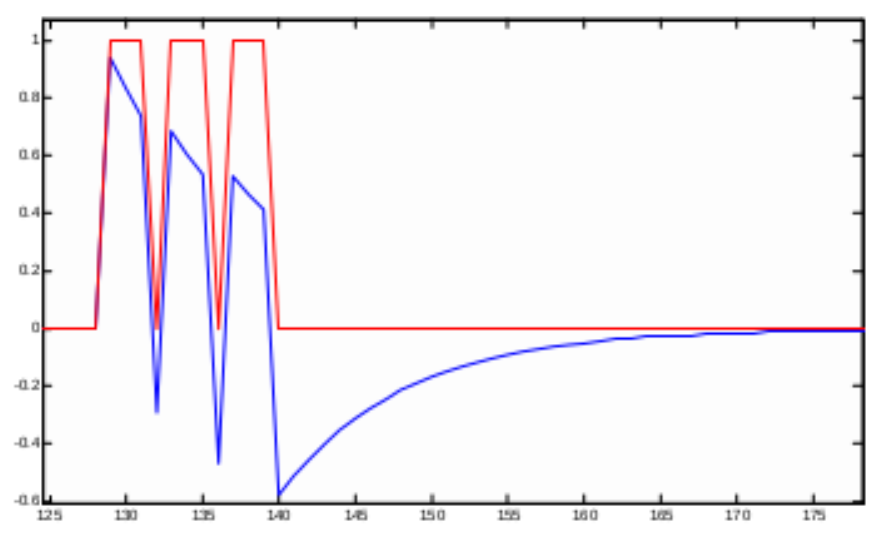}
\caption{\small The effect of a HPF in a square pulse (top panel) and in a train of square pulses (bottom panel).}
\label{fig:pulses}
\end{figure}

\begin{figure}[!htbp]
	\centering
	\subfloat[Blue - Real input signal; Orange - FEE output signal]{\includegraphics[width=0.5\textwidth]{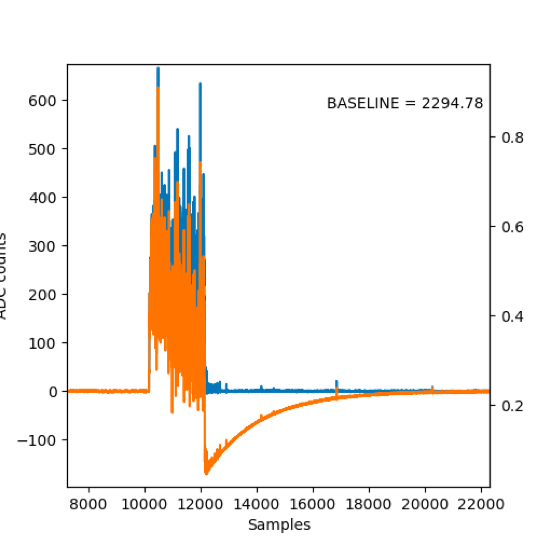}}
	\hfill
	\subfloat[Digital baseline reconstruction applied]{\includegraphics[width=0.5\textwidth]{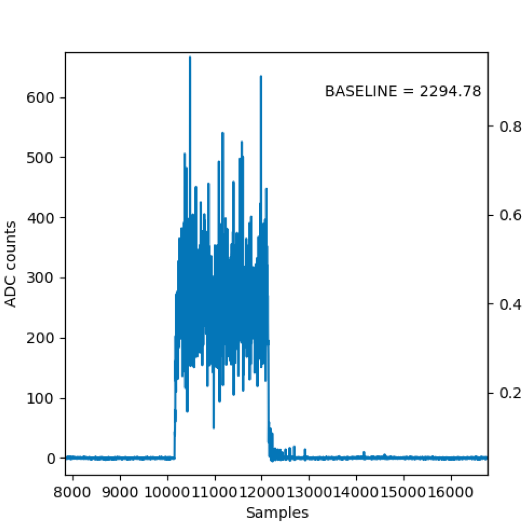}}
	\caption{\small Signal from 50us pulse. 25865 pe. Noise 0.74 LSB}
	\label{fig:graph_sig}
\end{figure} 

An AC coupling scheme creates, to first order, a high pass filter (HPF).  The effect of such a filter  
on a square pulse and on a train of squared pulses is shown in Figure \ref{fig:pulses}. The output pulse after the filter is the derivative of the input pulse, and is characterized by a null total area. Since the energy is proportional to the area of the input pulse, it follows that a deconvolution (or baseline restoration) algorithm must be applied to the output pulse in order to recover the input pulse area and thus measure the energy. Also, since the intrinsic energy resolution in \NEW\ is very good (of the order of 0.5 \% FWHM for point like particles), the error introduced by the deconvolution algorithm must be, at most, a fraction of a per mil. 

The energy plane front-end electronics and baseline restoration algorithm (BLR) has been described with detail elsewhere \cite{EnergyPlane2018}. In essence, the BLR algorithm is based on the implementation of the inverse function of a HPF. \Fig\ \ref{fig:graph_sig}  illustrates the action of the BLR algorithm plotting the output signal of a PMT (showing the characteristic negative swing introduced by the HPF) and the corrected signal after the base line restorarion. The estimated residual in the energy correction has been computed applying the deconvolution procedure to Monte Carlo signals and the effect has been quantified to be smaller than 0.3 \% FWHM for long signals (corresponding to large energies), and thus introducing an effect in the resolution smaller than the Fano factor \cite{EnergyPlane2018}.

%% file: src/tracking_plane.tex
\section{Tracking plane}
\label{sec.tp}

\begin{figure}[!htb]
\centering
\includegraphics[angle=0, width=0.47\textwidth]{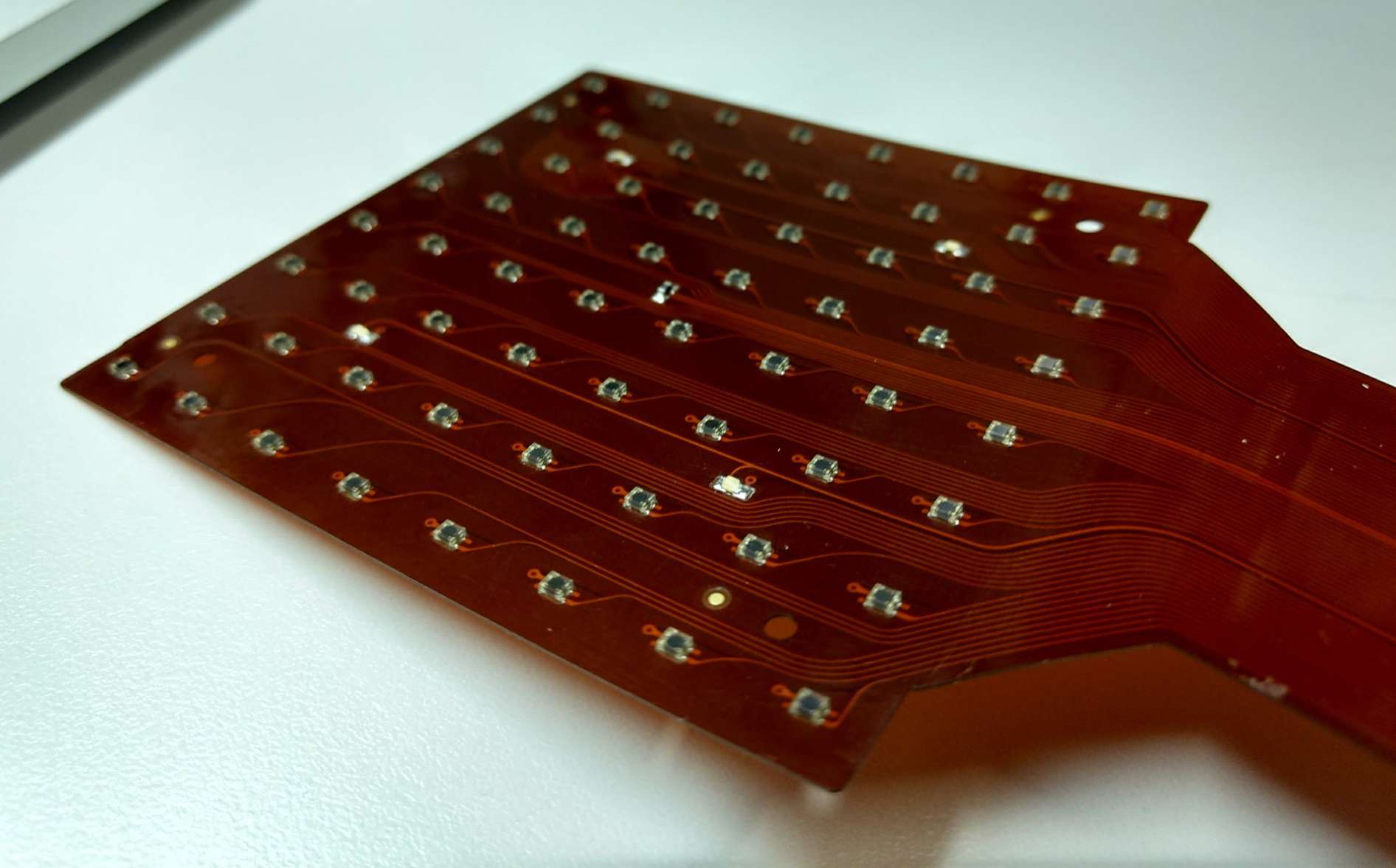}
\caption{\small A DICE board made of kapton, deploying 8 x 8 SiPMs.} 
\label{fig.dice}
\end{figure}

The tracking function in \NEW\ is performed by a plane holding a sparse matrix of SiPMs. The sensors active area have a size of  \NewSiPMSize\ and  are placed at a pitch of \NewSipmPitch. The tracking plane is placed \TrackingPlaneToAnode\ behind the end of the quartz plate that defines the anode with a total distance to the center of the EL region of \TrackingPlaneToEL.

The sensors are \NewSiPMSeries\ series model \NewSiPMModel\   with \NewSipmCell\ cell size and a dark count of less than \NewSipmDarkCount\ at room temperature. This implies that signals as low as 3 photo-electrons (pes) are already well above the dark count signal. The cell's size is sufficient to guarantee good linearity and the number of cells, 576 per sensor, combined with the PDE avoid saturation in the expected operating regime ($\sim$ \NewPhotoelectronsPerSiPM\ per SiPM). 

\begin{figure}[!htb]
\centering
\includegraphics[angle=0, width=0.47\textwidth]{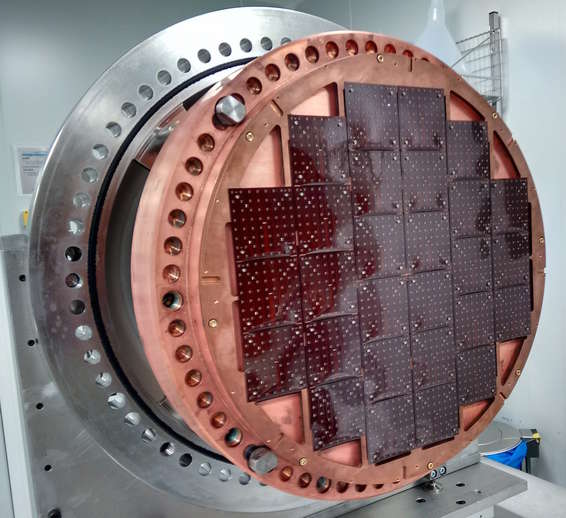}
\includegraphics[angle=0, width=0.47\textwidth]{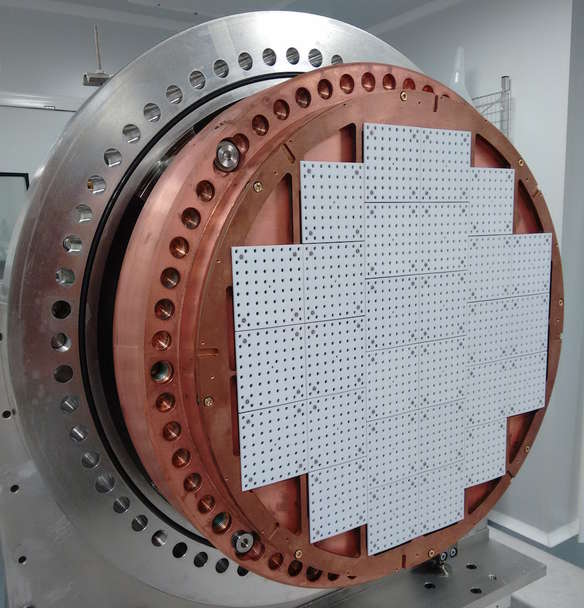}
\caption{\small Left: the 28 kapton DICE boards mounted on the tracking plane. Right: tracking plane upgraded with PTFE masks to improve the reflectivity in order to collect more light on the energy plane.} 
\label{fig:TP1}
\end{figure}

\begin{figure}[!htb]
\centering
\includegraphics[angle=0, width=1\textwidth]{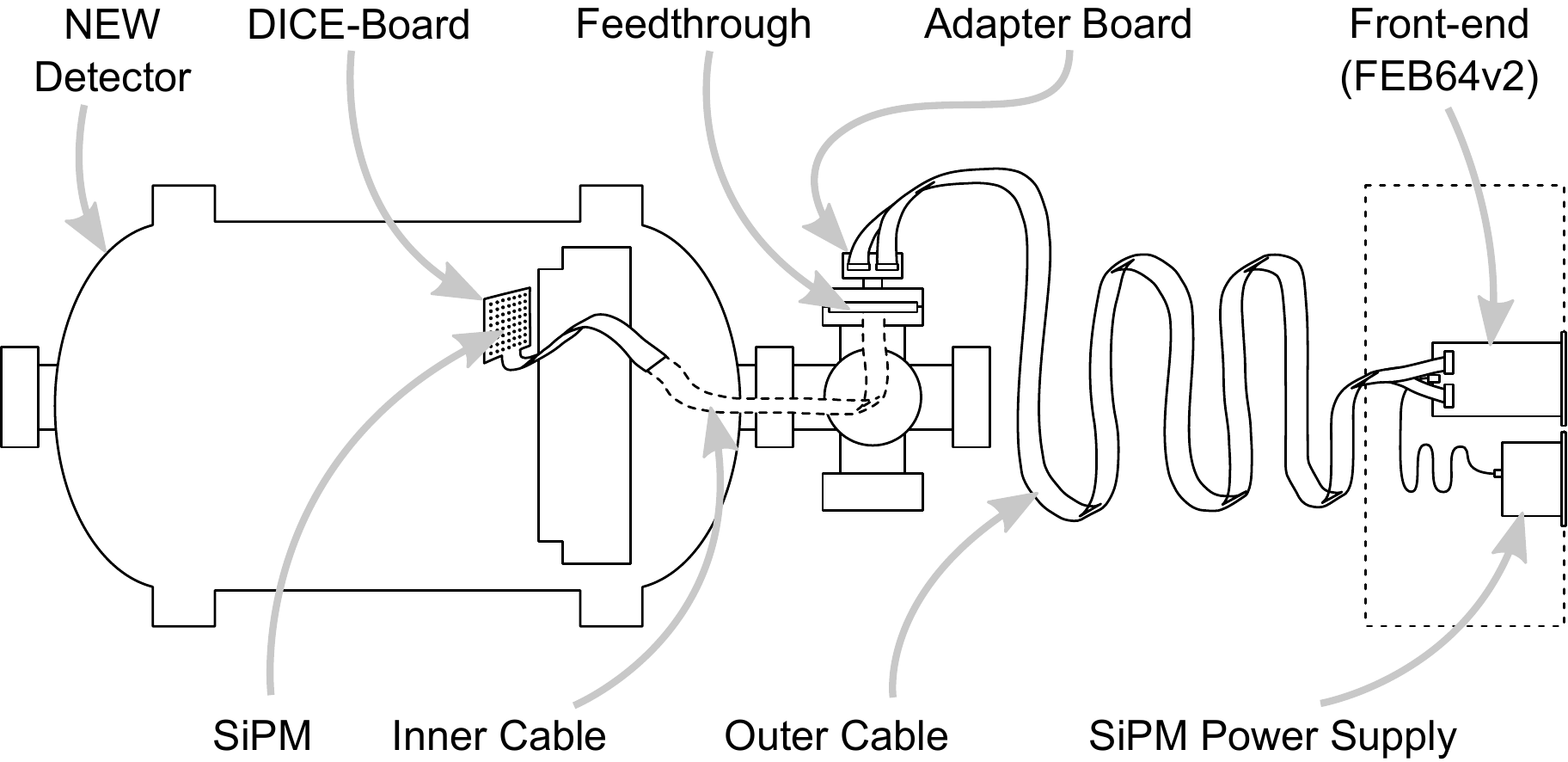}
\caption{\small \NEW\ Detector tracking plane scheme. The full signal chain from the silicon
photomultipliers to the front-end electronics is shown.} 
\label{fig:TP2}
\end{figure}

The SiPMs are distributed in \NewNumberOfBoards\ boards (DICE boards) with 
\NewNumberOfSiPMPerBoard\ pixels each (figure \ref{fig.dice}) for a total of \NewNumberOfSiPM\ sensors (figure \ref{fig:TP1}).  The DICE boards are mounted on a \NewTrackingPlaneEndCapThickness\ thick copper plate intended to shield against external radiation. The material used for the DICE boards is a low-radioactivity kapton printed circuit with a flexible pigtail that passes through the copper where it is connected to another kapton cable that brings the signal up to the feed-through (see figure \ref{fig:TP2}). Each DICE has a temperature sensor to monitor the temperature of the gas and SiPMs and also a blue LED to allow calibration of the PMTs in the opposite end of the detector. The \NEW\ tracking plane is currently the largest system deploying SiPMs as light pixels in the world.

\begin{figure}[!htb]
\centering
\includegraphics[angle=0, width=0.47\textwidth]{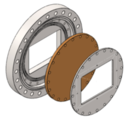}
\includegraphics[angle=0, width=0.47\textwidth]{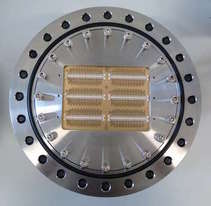}
\caption{\small Left: Feedthrough design (flange, FR4 PCB and stainless-steel reinforcement plate. The flange union with the PCB has a double elastomer gasket, which is butyl rubber o-rings. Right: finished feedthrough. For the sealing between the flange and the pipe we are using an elastomer gasket, but in the future it will be replaced by a metal gasket. Measured leak rate (Ar, 20 bar) was 10$^{-2}$ nmol/s, far below the target leak rate.} 
\label{fig:TP3}
\end{figure}

Taking into account that every SiPM has its own bias voltage wire, there are about \NewTrackingPlaneConnections\ electrical connections that must be passed from the inner volume of the pressure vessel to outside. A custom high-pressure, high-density feedthrough was developed based on a 3-layer 6-mm-thick FR4 PCB directly as a separation barrier between the pressurized xenon and the external air, taking advantage of the good mechanical characteristics of the FR4 to hold the pressure. The board stackup is designed with 3 copper layers and blind vias (drills) misaligned and filled with vacuum epoxy, so there is not a direct path for a gas leak. The feedthrough (see figure \ref{fig:TP3}) includes connectors for six DICE-Boards. Notice that the FR4 feedthroughs are shielded from the detector by the thick copper plate. 

Inner kapton cables are not a cost-effective solution as external cables. The best option was a commercial 4 meter 51-wire 0.5-mm-pitch ribbon cable (reference PS2829AA4000S from Parlex) embedded in flexible polyester substrate. It equips surface mount connectors at the ends (DF9B-51S-1V from Hirose).

\subsection{SiPM front-end electronics}
\begin{figure}[!htb]
\centering
\includegraphics[angle=0, width=1\textwidth]{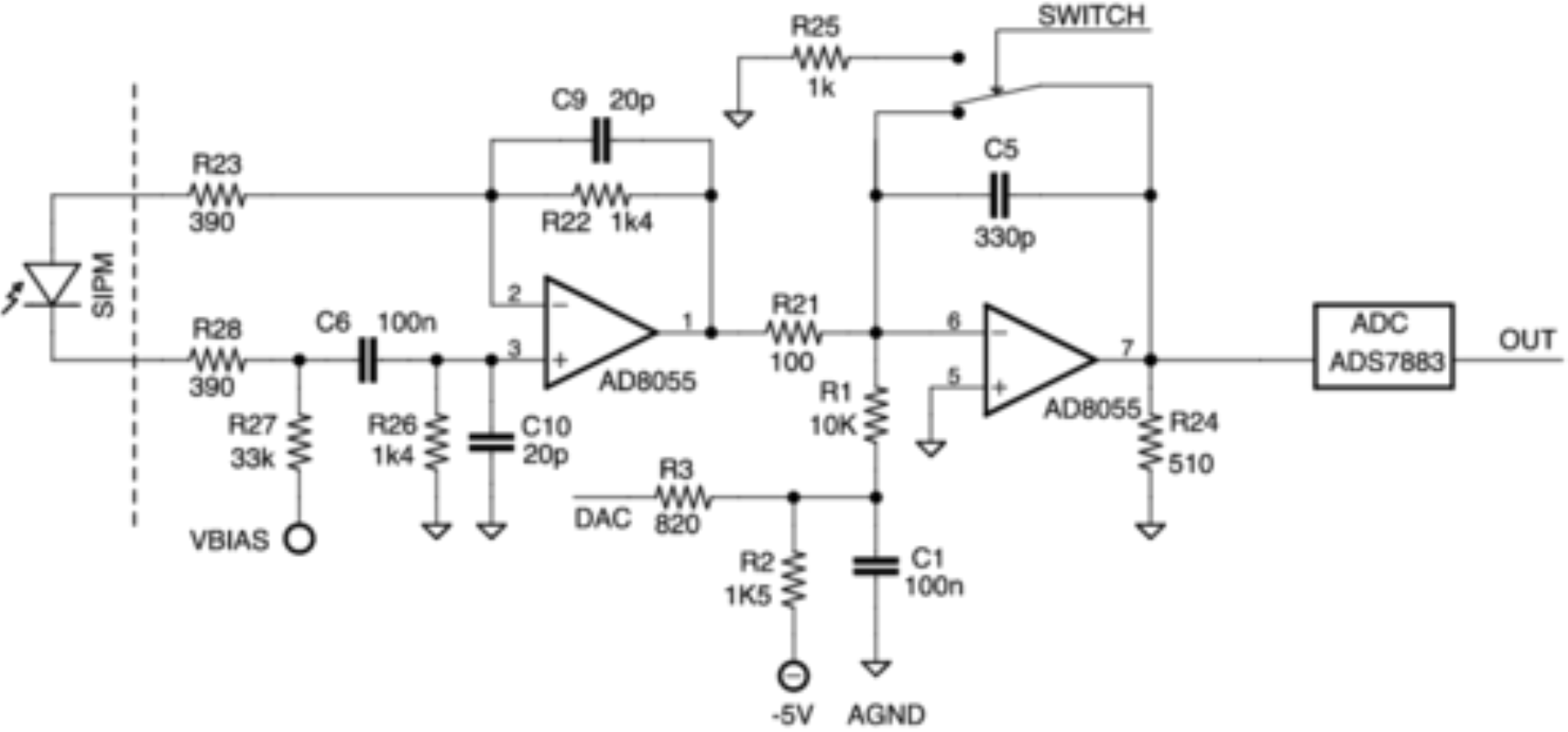}
\caption{\small Differential SiPM front-end channel.} 
\label{fig:SFE1}
\end{figure}

The time resolution requirement for the tracking plane was set to $1 \mu s$ for an electron drift velocity of 1mm/$\mathrm{\mu}$s. This allows simpler (and cheaper) solutions for the front-end electronics other than the conventional high-speed waveform sampling scheme.  Specifically, an approach based on a gated integrator has been chosen for \NEW. In this scheme, a differential-input transimpedance amplifier is followed by an integrator. After a  $1 \mu s$ charge accumulation time, a 12-bit 1-MHz ADC samples the integrated value. A switch discharges the integrating capacitor in approx. 20 ns and opens to allow a new  $1 \mu s$ integrating period (actually 980 ns). A DAC-driven offset compensation circuit sets the channel baseline to the desired value. ADC, DAC and switch are controlled by an FPGA. Figure \ref{fig:SFE1} shows the described front-end circuit with the actual component values. The front-end electronics of the tracking plane have been described in detail in \cite{Rodriguez:2015upa}.

%% file: src/data_acquistion.tex
\section{DAQ and trigger}
\label{sec.daq}

\subsection{DAQ}
\label{subsec:DAQ}
The DAQ software chosen for \NEW\ (and \NEXT) is DATE, a framework developed by the ALICE experiment at CERN\cite{Carena:2014wda}. The main reason for this choice has been the excellent scalability of the system and the robust support, given the wide user's base. 

Regarding DAQ hardware, the NEXT experiment, together with the RD51 Collaboration at CERN, and IFIN-HH Bucharest, co-developed SRS, the Scalable Readout System in 2011 \cite{Martoiu:2013aca}. SRS was intended as a general purpose multi-channel readout solution for a wide range of detector types and detector complexities. The first SRS version relied on the cost-effective 6U Eurocard mechanics. A generic FPGA-based readout card (the FEC card \cite{Toledo:2011}) together with an edge-monted front-end specific add-in card form a 6Ux200 mm Eurocard module. Several add-in cards were developed to read out popular ASICs in the RD51 community, to digitize analog channels (like PMTs) and to interface digital front ends. 

A second SRS version based on ATCA for enhanced reliability was released in 2014 and was adopted for the \NEW\ DAQ system. An ATCA chassis with redundant power supplies and cooling houses the DAQ blades, consisting each of two interconnected FPGAs, two mezzanine slots, two event buffers and I/O interfaces (several Gigabit Ethernet links and NIM trigger signals). As a result, an ATCA DAQ blade equips the resources of two ``classic" SRS FEC modules.

\subsubsection{ATCA-SRS, the \NEW\ Readout and DAQ System}
\label{subsubsec:SRS}
The \NEW\ DAQ system is divided into three partitions: the energy plane (\NewNumberOfPMT\  PMTs), the tracking plane (\NewNumberOfSiPM\ SiPMs grouped in \NewNumberOfBoards\ front-end read out boards) and the trigger partition. As discussed in \cite{Esteve:2016}, PMT signals are digitized at \NewPMTDigiSpeed, \NewPMTADCBits bit. Once digitized and framed, this produces approximately \NewPMTDataFlow\ for the \NewNumberOfPMT\ PMTs at a \NewTriggerRateBase\ trigger rate. Regarding the tracking plane partition, SiPM data once integrated, digitized (12 bit, 1 MHz) and framed results in approx. 1 Mb of data per event per front-end board. For  \NewNumberOfBoards\ front-end boards, 1 ms events and a \NewTriggerRateBase\ trigger rate, this partition generates \NewSiPMDataFlow\ in raw data mode. The trigger partition generates a tiny amount of data.

\begin{figure}[!htb]
\centering
\includegraphics[angle=0, width=1\textwidth]{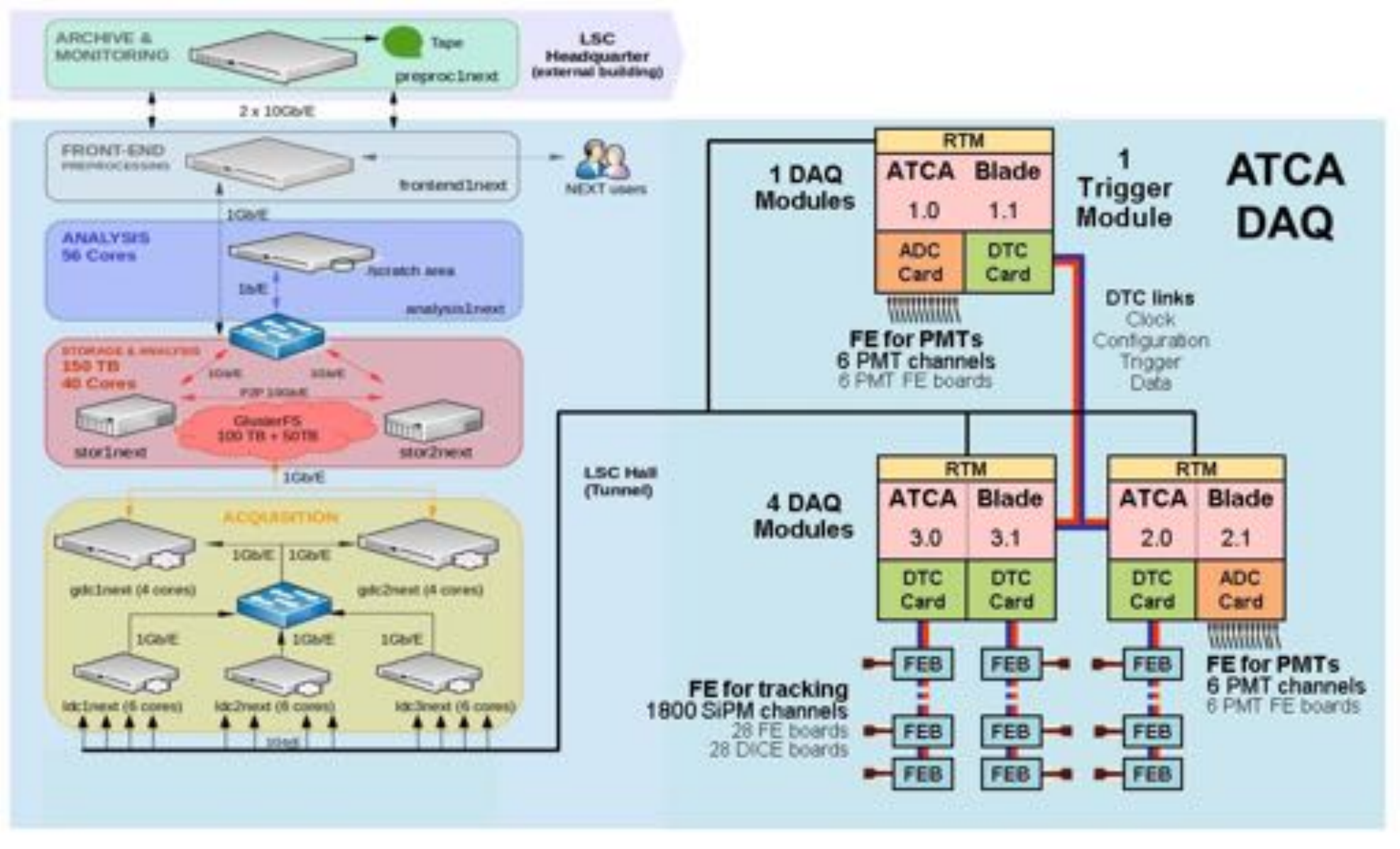}
\caption{\small \NEW\ DAQ and storage architecture.} 
\label{fig:DAQ}
\end{figure}

Figure \ref{fig:DAQ} shows the DAQ architecture. Three halves of ATCA blades (3.0, 3.1 and 2.0 in figure \ref{fig:DAQ}) equipped with LVDS interface mezzanines (named DTC cards, for Data, Trigger and Control) are required to read out the \NewNumberOfBoards\ SiPM front-end boards. Just one half of ATCA blade with an ADC mezzanine suffices to read out the \NewNumberOfPMT\ PMTs in the energy partition, though two mezzanines are used for reduced noise. This accounts for two and a half ATCA blades.

An additional half of ATCA blade (1.1 in figure \ref{fig:DAQ})  is used as trigger module, receiving trigger candidate data from the PMT data, applying the configurable trigger algorithm described in section \ref{subsec:Trgg} and distributing timestamped triggers to the other ATCA blades. The trigger module is also responsible for sending configuration and synchronization messages to the other FEC blades.

\subsubsection{DAQ computers}
\label{subsubsec:computers}
Data are sent from the SRS-ATCA blades to the DAQ PCs using optical Gigabit Ethernet links. DATE defines a hierarchy in the DAQ computers in which local data concentrator PCs (LDCs) receive data from the SRS-ATCA blades and send larger, merged sub-events to a layer of global data concetrator PCs (GDCs). LCDs and GDCs are nodes in a Gigabit Ethernet network. Finally, GDCs send complete events to the final storage.

In \NEW\, three LDCs and two GDC suffice to handle the event data load. In order to balance the load, a GDC receives only even or odd event number data. A first storage consists of two PCs with a total storage capacity of 150 TByte on swappable hard disks. Events are filtered before sent to a final storage in tape.

\subsection{Trigger}
\label{subsec:Trgg}

The trigger in \NEW\ is an evolution of the trigger developed for the DEMO prototype \cite{Esteve:2012hy} that adds ample flexibility on the event selection. This is needed since the expected \st\ signals of MeV electrons in a \HPXe\ are very diverse given the large variety of possible orientations of these electrons.

In the case of \NEW\ we have implemented a larger buffer, up to \NewMaxTriggerBuffer, corresponding to roughly three times the length of the NEXT-100 detector (thus 6 times the length of \NEW). In order to minimize the dead time due to such a large buffer, a system of double buffer has been developed. In calibration data both triggers are used for all events, while in a physics run it is possible to define \textit{interesting events}, usually events near \Qbb, that have priority in the usage of these buffers reducing the likelihood of losing one of these events. For trigger rates as high as 1 Hz of interesting events, the dead time is negligible (<0.1\%) using zero suppressed data.

%


%% file: src/gas_system.tex
\section{Gas system}
\label{sec.gas}

\subsection{Overview of the GS}

\begin{figure}[!htb]
\centering
\includegraphics[angle=0, width=0.99\textwidth]{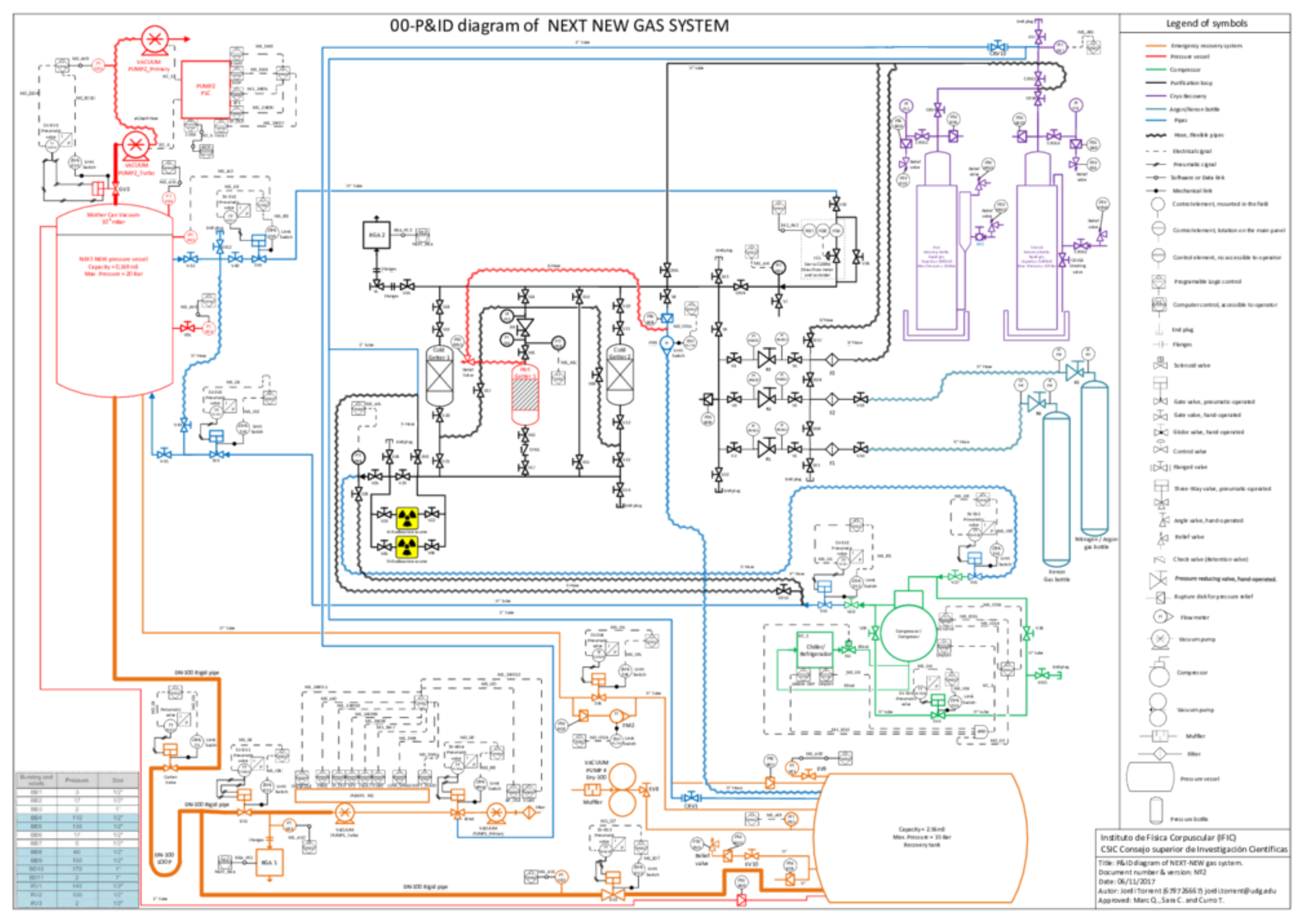}
\caption{Diagram of the gas system for the \NEW\ detector.} 
\label{fig:gas_system}
\end{figure}

Achieving a long electron lifetime (compared with the drift length of the detector) is a must for all noble gas or liquid TPC experiments, and in particular for NEXT, given its excellent energy resolution. In the case of NEXT-100, the target lifetime is of the order of 5 ms (the detector drift length is 1 ms). This requires controlling the impurities (in particular oxygen) at the level of one part per billion (1 ppb) \cite{Herzenberg:O2, HUK1988107}, that can only be achieved by careful selection of the TPC materials in contact with the gas and by continuous circulation and purification of the gas. 

The role of the \NEW\ and NEXT-100 gas system (GS hereafter) is the purification of the xenon, achieved by circulating the gas through chemical purifiers (getters). In the GS, pressure must be maintained stable at nominal values at all times, and the system must react quickly and automatically to significant changes indicating an unexpected over-pressure condition or a leak. Furthermore, continuous monitoring allows the detection of small variations in pressure due to temperature oscillations (also measured across the system) or small leaks. 

It follows that the GS must include at least the following elements: a) an inlet, where the gas is fed into the system; b) recirculation machinery (e.g., a compressor), used to keep the pressure of the system at the nominal value and to circulate the gas; c) purification devices (getters); d) an emergency recovery tank (also called expansion tank) capable to recover quickly the gas if need arises; e) a cryo-recovery bottle to recover the gas in normal operation conditions; f) sensor devices, to monitor the state of the system and trigger automatic actions (such as an emergency recovery) if needed; g) vacumm system to evacuate the lines before their filling with xenon.

Figure \ref{fig:gas_system} shows a functional diagram of the GS which can be described in terms of four subsystems, namely: a) the pressure system (PS), encompassing gas recirculation and purification; b) the vacuum system (VS), which includes all the vacuum lines and pumps connected to the different parts of the system; c) the emergency recovery system (ERS), whose main part is the expansion tank and d) cryo-recovery system (CRS), including the the cryogenic recovery bottle and all the lines connecting it to the rest of the system. The four systems are all interconnected. 
 
\subsection{Pressure system}

\begin{figure}[!htb]
\centering
\includegraphics[angle=0, width=0.8\textwidth]{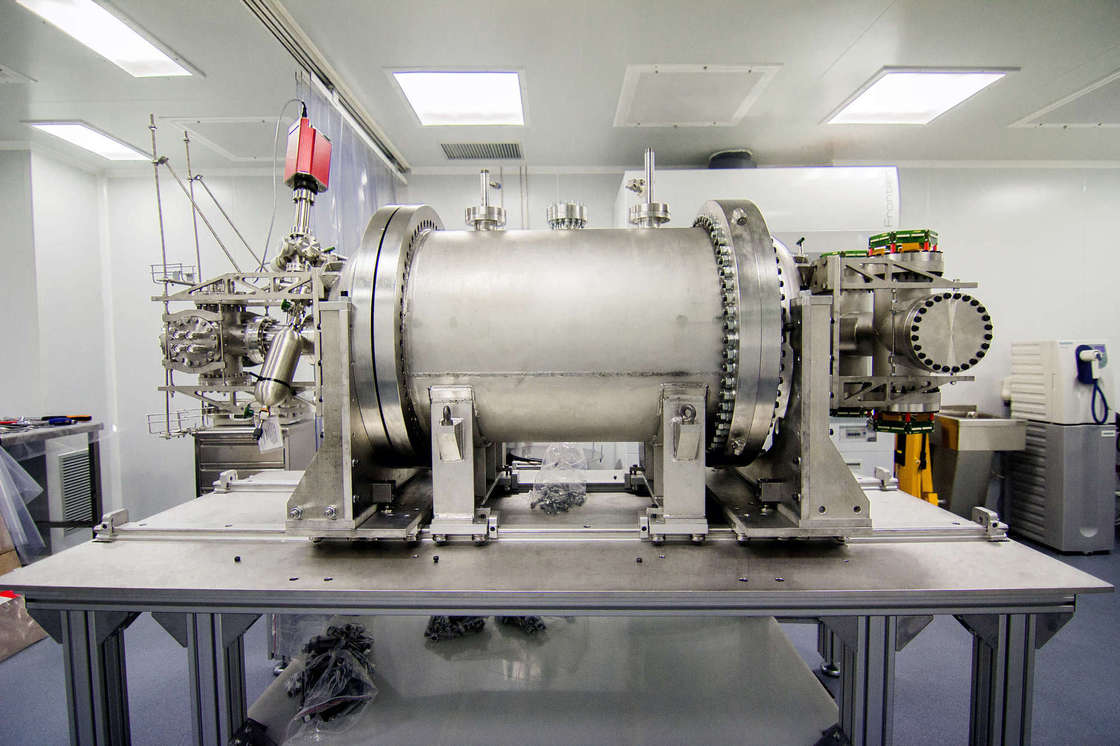}
\caption{Image of the \NEW\ vessel in the LSC clean room while detector assembly. } 
\label{fig:vessel}
\end{figure}

\begin{figure}[!htb]
\centering
\includegraphics[angle=0, width=0.8\textwidth]{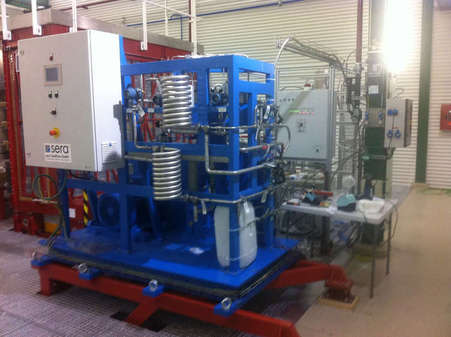}
\caption{The \NEW\/NEXT-100 compressor, manufactured by SERA.} 
\label{fig:compressor}
\end{figure}

The total volume of the PS is \NewPSVolume, shared between the main pressure vessel (\NewMainPSVolume),  the gas loop (\NewGasLoopVolume) and the compressor (\NewGasLoopVolume).
The pressure vessel is certified to operate at \NewMaxPressure\ while the hot getter can only tolerate \NewGetterPressure, thus imposing the use of a pressure regulator at the getters inlet. 

The pressure vessel was built using stainless steel 316 Titanium alloy (TP316Ti). The central body of the vessel (Fig. \ref{fig:vessel}) has a total length of \NewPressureVesselLength\  an internal diameter of \NewPressureVesselDiameter, and wall thickness (in the barrel region) of  \NewPressureVesselBarrelThickness. The steel alloy was selected by its mechanical strength and its remarkable radiopurity.  

The gas loop has two \NewColdGetters\ ambient temperature getters \cite{saes:coldgetter} (cold getters) and one \NewHotGetters\ hot getter \cite{saes:hotgetter}.
The three getters can be operated in parallel to allow large flow operations. The normal mode of operation is to circulate only through the hot getter, given the large radon contamination introduced by the cold getters. Indeed, recirculation through the hot getter only is a must in a physics run. However, during commissioning, circulation through the cold getters (or both cold and hot getters) allows a faster purification of the gas and is an acceptable strategy provided sufficient time is allowed after closing the cold getters for the radon to decay.

The compressor, shown in figure \ref{fig:compressor} was manufactured by SERA \cite{sera:compressor}. The inlet takes gas at a pressure between \NewCompressorMinimumInletPressure\ and \NewCompressorMaximumInletPressure. The maximum outlet pressure is  \NewCompressorMaximumOutletPressure. The total leak rate of the compressor has been measure to be smaller than \CompressorLeakRatePerYear. The compressor has a triple diaphragm system that prevents catastrophic loses of xenon and gas contamination in case of an abrupt diaphragm rupture.

The flow direction inside the active volume of the detector is from the anode to the cathode. The gas enters just behind the tracking plane copper plate (See sec. \ref{sec.tp}) and exits through a VCR 1/2" port just after the cathode. With this configuration, the clean gas enters directly into the amplification region.

\subsection{Cryo-recovery system}

\begin{figure}[!htb]
\centering
\includegraphics[angle=0, width=0.4\textwidth]{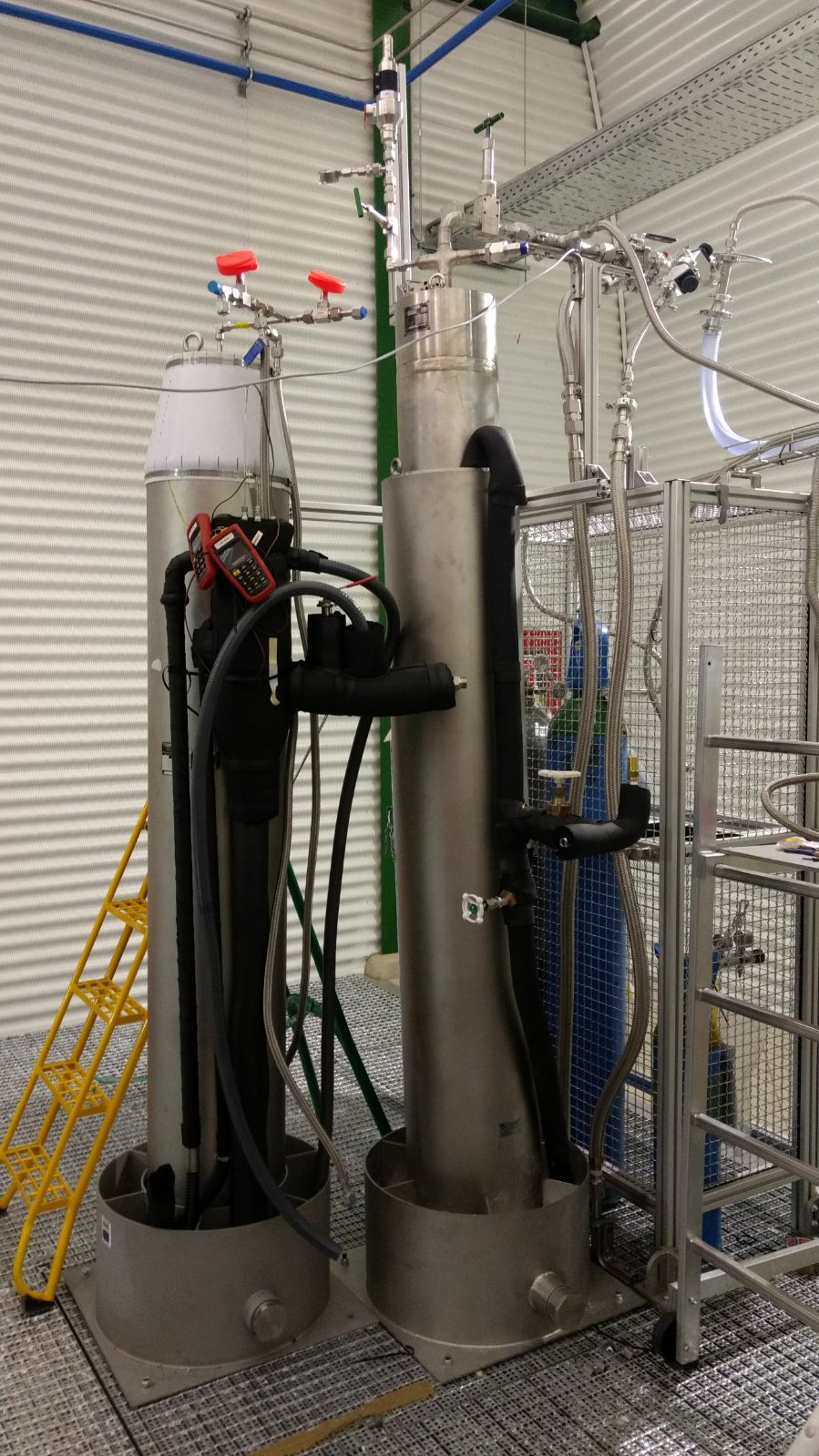}
\includegraphics[angle=0, width=0.4\textwidth]{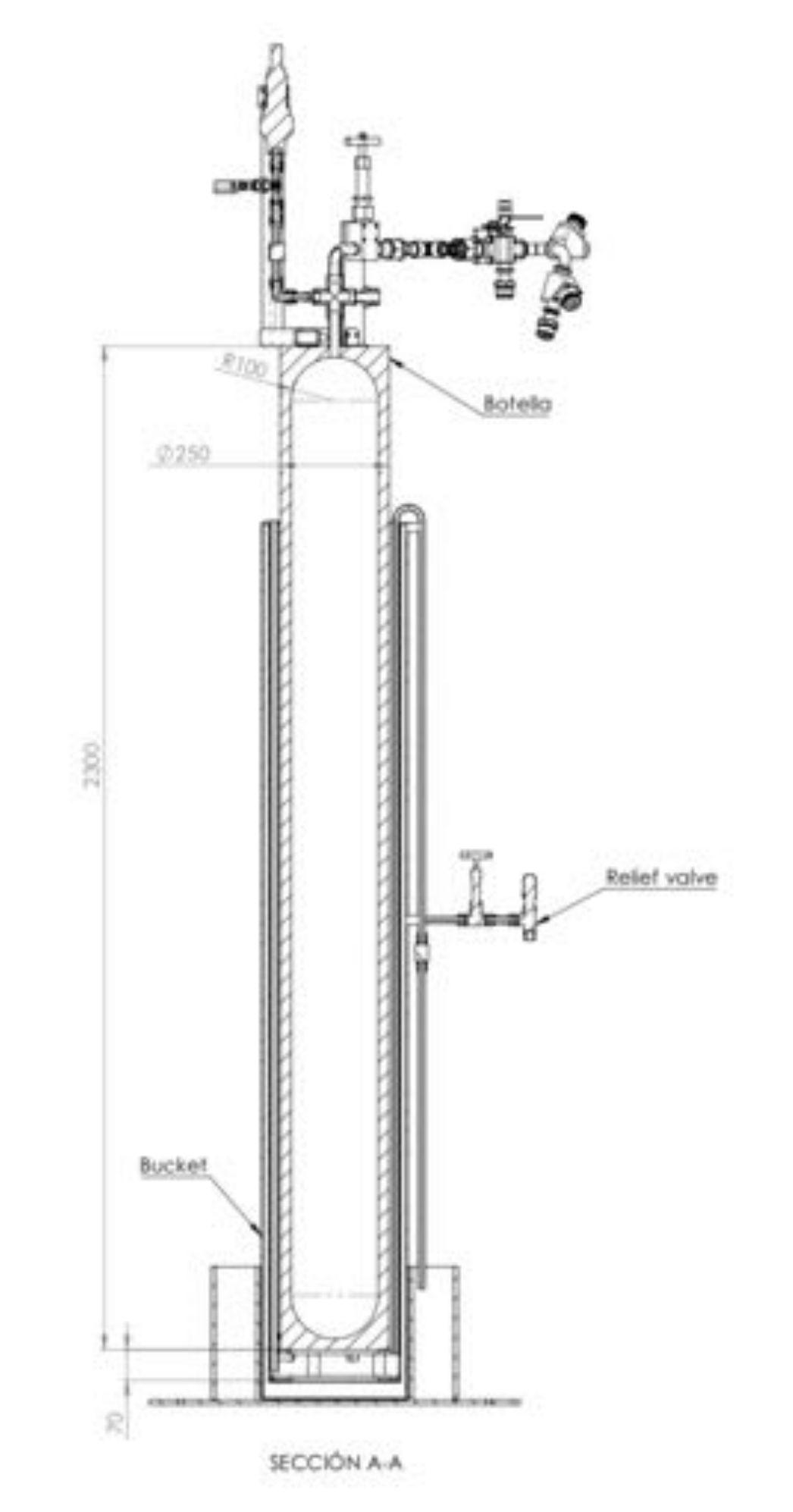}
\caption{Left: A picture of the two NEXT cryo-bottles. Right: A drawing showing the design of the second bottle (the first one is similar). }
\label{fig:cryo34}
\end{figure}

Recovery of the xenon under normal operation conditions is achieved by connecting two
cryo-bottles (figure \ref{fig:cryo34} left) to the recirculation loop and to the expansion tank (so that one can recover the xenon stored there in the event of an emergency evacuation).  One of the bottles is used to recover normal xenon, while the other will be used to recover enriched xenon.  Each bottle is placed inside its own cryo-bucket. To recover the gas, one of the two bottles is cooled down to \LINNormalTemperature\ using liquid nitrogen, thus creating a pressure gradient that results in xenon gas flowing into the cold vessel where it solidifies, so that the pressure does not increase while the gas is being recovered. The bottle is then closed and let warm up as the nitrogen evaporates.

The cryo-bottles consist of two cylinders: the external cylinder (bucket) has double walls to provide thermal isolation and its separated from the internal cylinder (bottle), used to recover the gas, by a few millimeters air gap (figure \ref{fig:cryo34} right). Both bottles have similar capacity (\NewFirstBottleVolume\ and \NewSecondBottleVolume). They are designed to be able to recover the full amount of gas circulating in NEXT-100 (e.g., \NewSecondBottleXenonMass\ of enriched xenon in the case of the second bottle).  

%
%
%
%

Two cryogenic recovery operations with xenon were successfully carried out in December 2016. The first operation recovered the xenon from the gas system pipes and the vessel. The second, from the expansion tank. In both bases the remaining residual pressure was very small, corresponding to a few tens of grams left in the system. 

\subsection{Emergency system}

\begin{figure}[!htb]
\centering
\includegraphics[angle=0, width=0.9\textwidth]{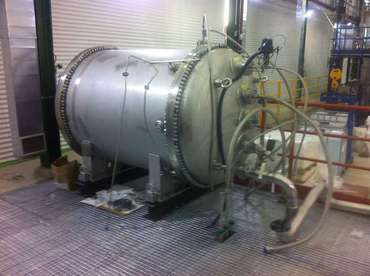}
\caption{The NEXT-100 pressure vessel is used as expansion tank during the operation of the \NEW\ detector. A new expansion tank with a volume ten times larger will be installed for the operation of the NEXT-100 detector. }
\label{fig.tank}
\end{figure}

The role of the emergency system is to recover quickly the gas into an expansion tank if need arises (e.g, a large leak or an over-pressure condition). For that purpose, an expansion volume one order of magnitude larger than the xenon volume contained in the gas system permits the recovery of the gas in seconds. 

Figure \ref{fig.tank} shows a picture of the NEXT-100 pressure vessel which is being used as the expansion tank for the operation of the \NEW\ detector. The tank has a volume of \NewExpansionTankVolume\ and is connected to the main vessel through a CF 100 sealed with a Carten valve and a 1/2" line with a pneumatic actuated valve in parallel with a bursting disk. The pressure in the vessel is being continuously monitored and a dedicated slow control is responsible for activating the recovery system (opening the Carten and the pneumatic valve) in case of an unexpected pressure drop. The case of over-pressure is covered by the bursting disk which breaks if the pressure exceeds a tolerable level, thus triggering the recovery. 




\subsection{Vacuum system}

The role of the vacuum system is to reduce the contamination of other gases and/or volatile substances (e.g, air, nitrogen, oxygen, methane, oils, etc.) that may be trapped in the detector
, or the various parts (pipes, valves, etc) of the gas system, before adding xenon. Vacuum-pumping of the system also reduces the outgassing of materials in the pressure vessel, reducing considerably the time needed to purify the gas. 
The system consists of a series of pumps (see table \ref{tab:Pumps}) connected to different parts (vessel, circulation, recovery) of the system. 


\input{src/TablePumps}



%% file: src/TablePumps.tex
\begin{table}[ht!]
\centering
\begin{tabular}{ c c c c }
System & Pump & \parbox[t]{1.8cm}{Nominal capacity \\ (m$^3$/h)} & \parbox[t]{2cm}{Ultimate pressure \\(mbar)} \\ \hline\hline
\multirow{2}{*}{Pressure vessel} & Scroll-IDP15  & 12.8  & 13$\cdot$10$^{-3}$ \\
 & Agilent twistoor 304 FS  & 958 & <5 x 10$^{-10}$ \\ \hline
\multirow{3}{*}{Emergency recovery} & PVR DRY 100 & 120 & 150  \\
 & Scroll-IDP15  & 12.8  & 13$\cdot$10$^{-3}$ \\
 & Agilent twistoor 304 FS  & 958 & <5 x 10$^{-10}$ \\ \hline
\multirow{2}{*}{Gas loop} & nxds20i Edwards & 22 & 0.03 \\
 & Turbo nEXT300D & 46 &  <5 x 10$^{-10}$ \\ 
\end{tabular}
\caption{Vacuum pump characteristics grouped by systems. Turbo and scroll pumps used in NEW vessel are also used for the emergency recovery tank. }
\label{tab:Pumps}
\end{table}

%% file: src/sensor_calibration.tex
\section{Sensor calibration}
\label{sec.calib}

\subsection{Calibration of the energy plane}

\begin{figure}[!htb]
  \begin{center}
    \includegraphics[width=0.49\textwidth]{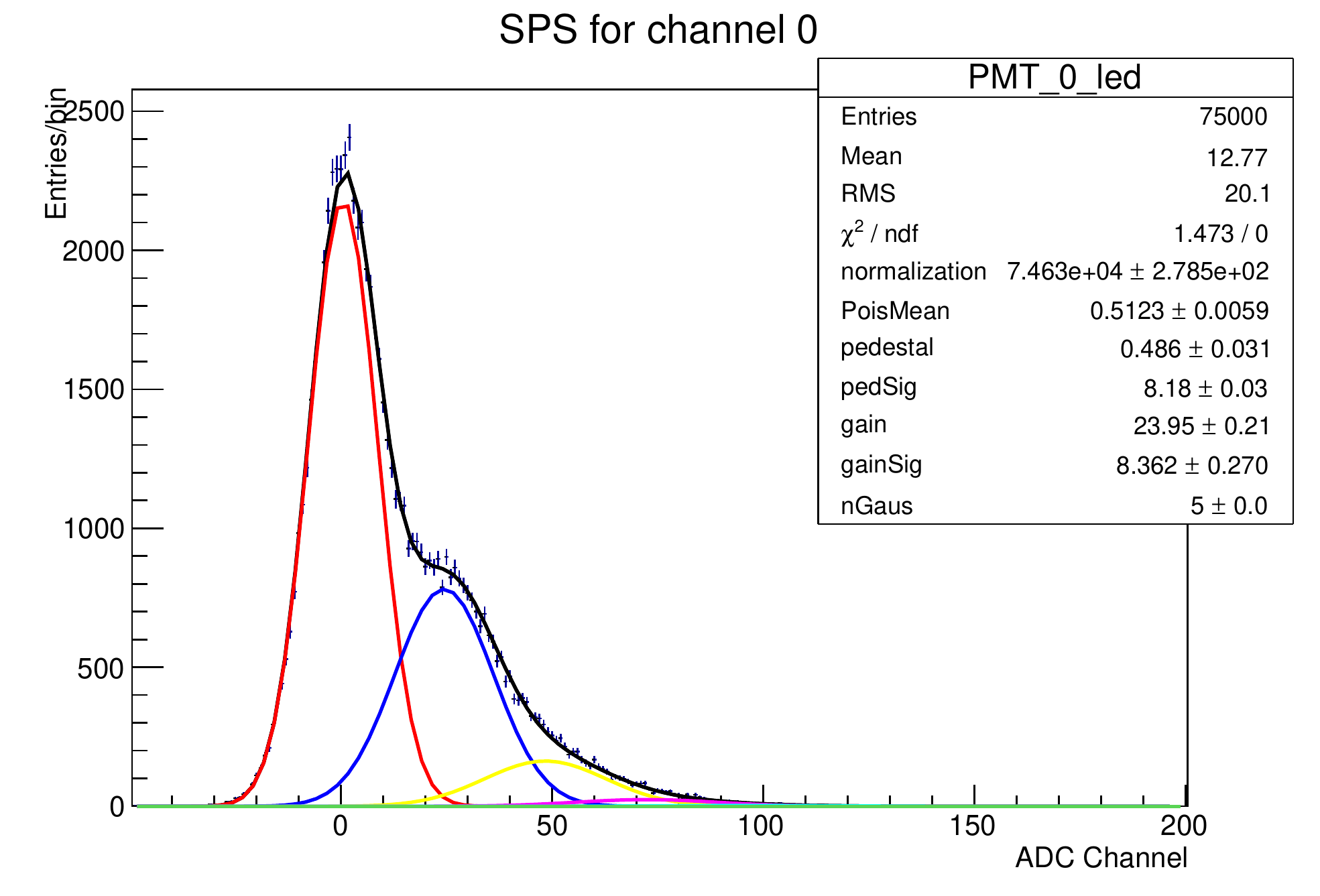}
    \includegraphics[width=0.49\textwidth]{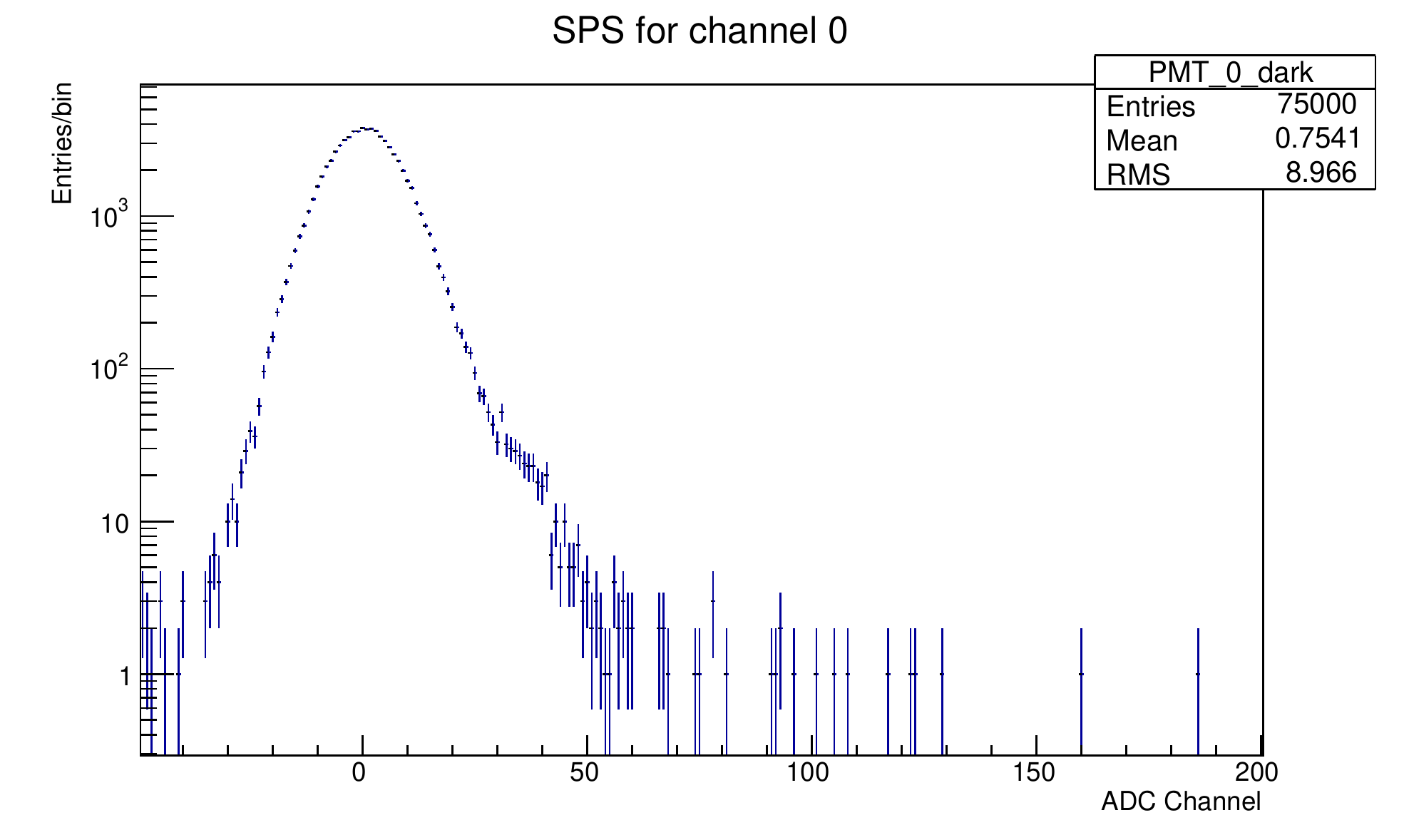}
  \end{center}
  \caption{Standard PMT calibration spectra. Left: SPE spectrum with fitted Gaussians. Right: ``dark'' spectrum showing dominant electronic noise with contribution form dark current.}
  \label{fig:specEx}
\end{figure}

The main role of the energy plane in \NEW\ is the measurement of the event energy. This requires a precise equalization of the PMTs comprising the energy plane, obtained through regular LED calibrations. The LEDs are located in the tracking plane, facing the PMTs and is pulsed every
\NewEnergyPlaneLedPulse. Readout is then triggered automatically to accumulate several thousands of events, divided in several runs where the LED light is adjusted to various levels that approximate the single photon regime (spe). Checking the uniformity of results in the various runs, permits to ensure that the gain and charge resolution determined by the procedure are not affected by the light levels used for the calibration.

The data are then subject to low level processing which carry the baseline restoration of the waveform, remove the pedestals and integrate the signal in the regions where the pulses are expected as well as in regions about  \SI{2}{\mu s} before the pulse. This results both spectra with contributions from zero to $n$~ photoelectrons and in spectra of electronic noise with a small contribution from dark noise, respectively, (figure \ref{fig:specEx}) which can be used to extract key parameters.


\begin{figure}[!htb]
  \begin{center}
    \includegraphics[width=0.49\textwidth]{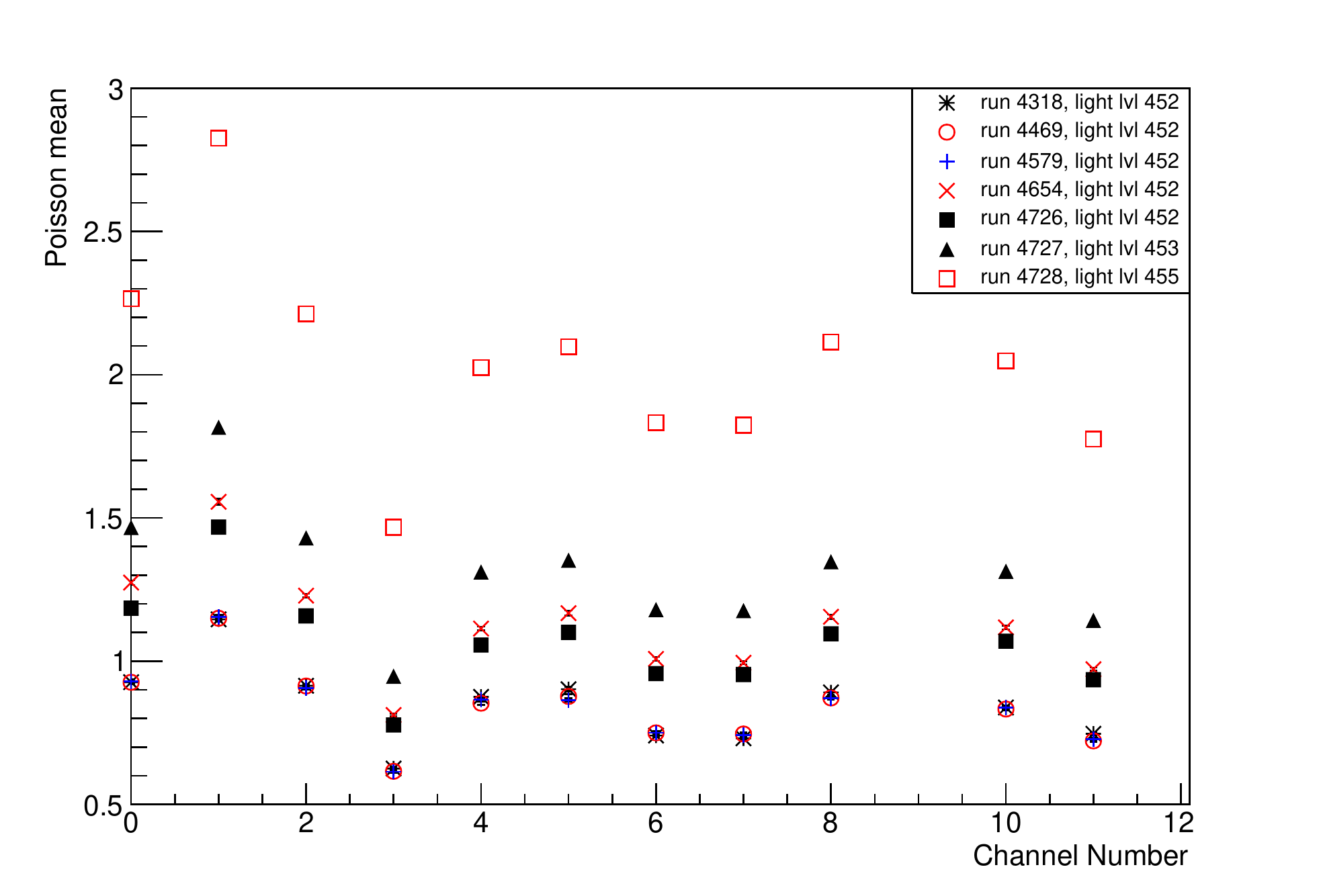}
    \includegraphics[width=0.49\textwidth]{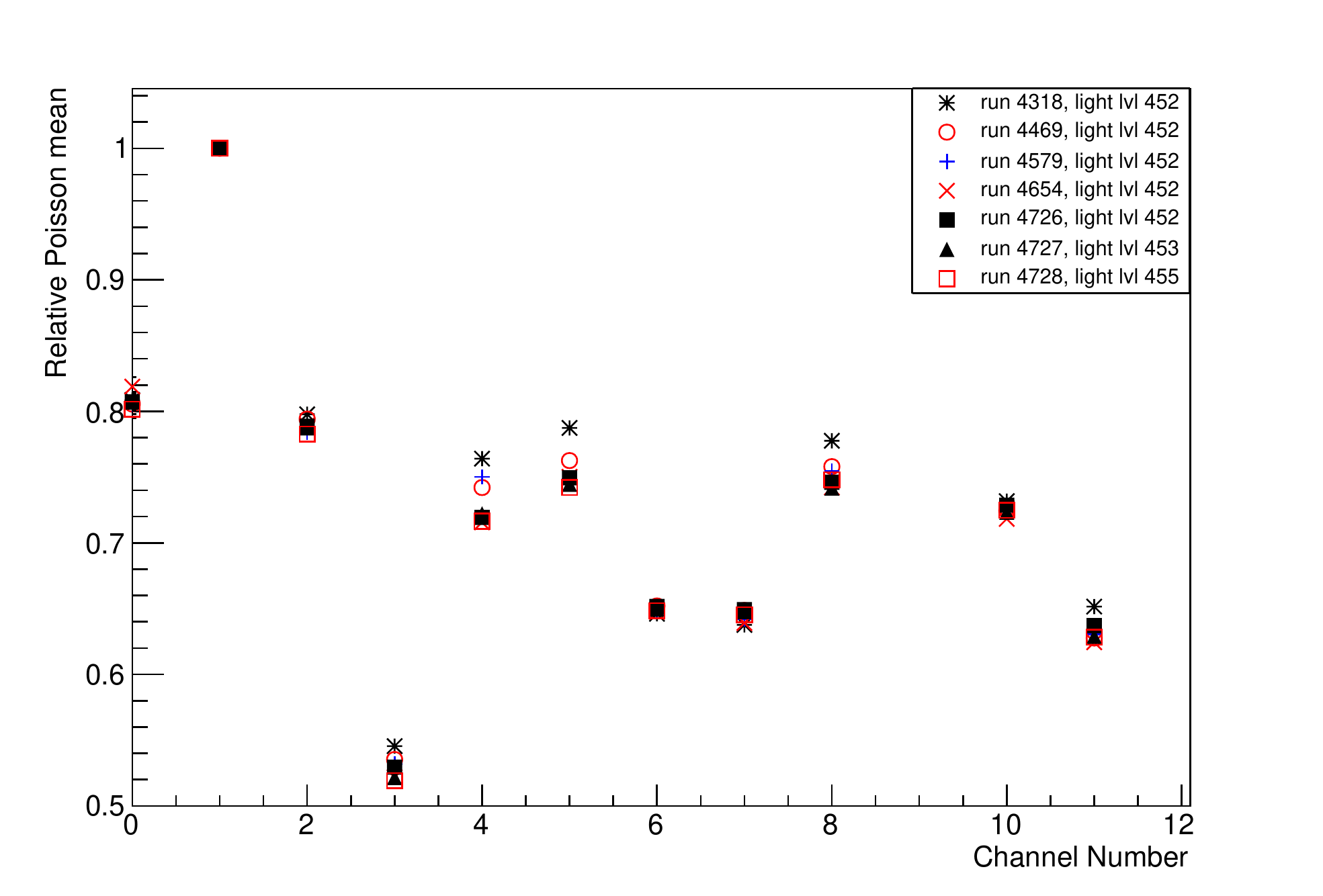}
  \end{center}
  \caption{Left: Poisson mean values from various PMT calibration runs; Right: the same values normalised to that for the channel which sees most light.}
  \label{fig:PoisM}
\end{figure}

\begin{figure}[!htb]
  \begin{center}
    \includegraphics[width=0.49\textwidth]{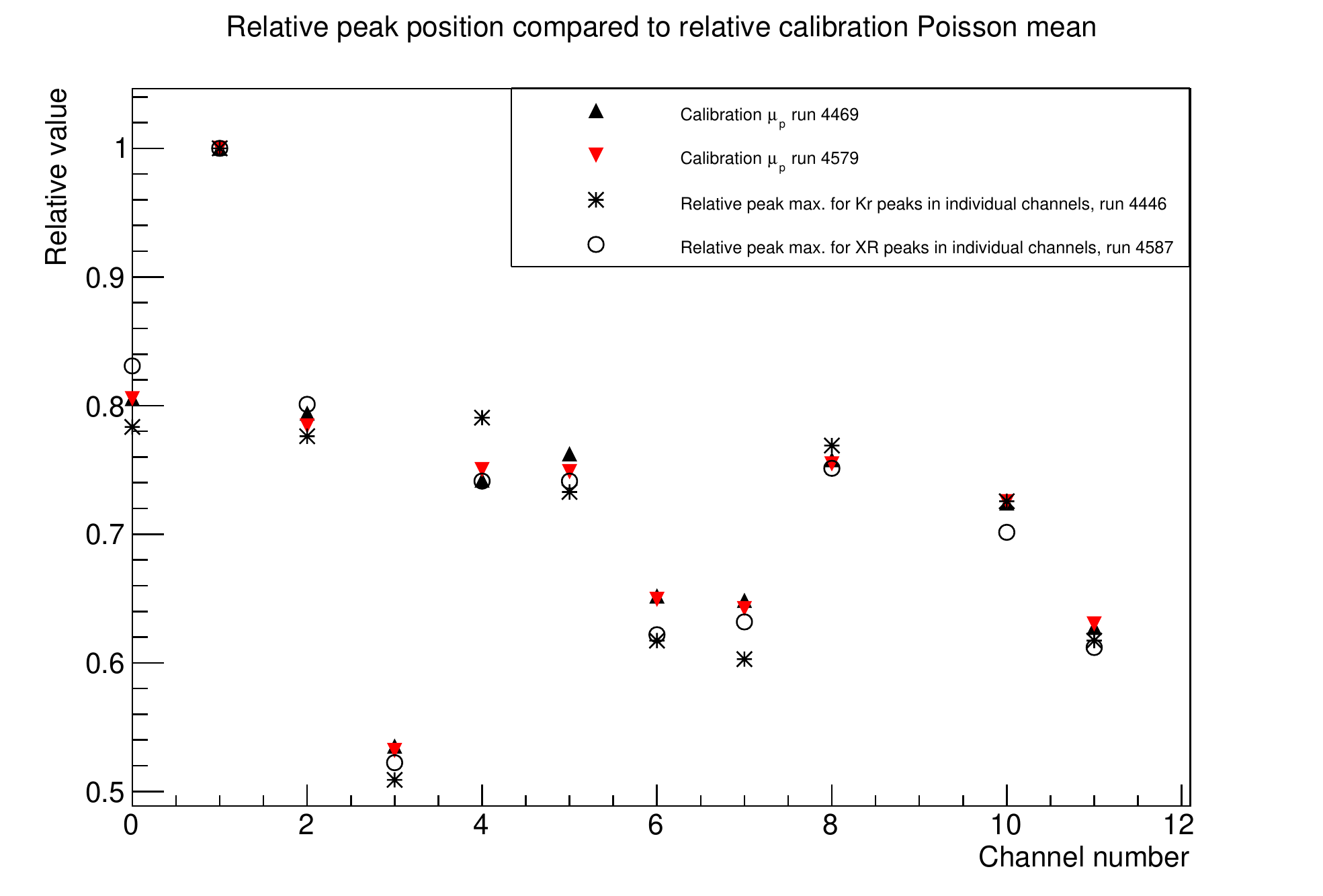}
    \includegraphics[width=0.4\textwidth]{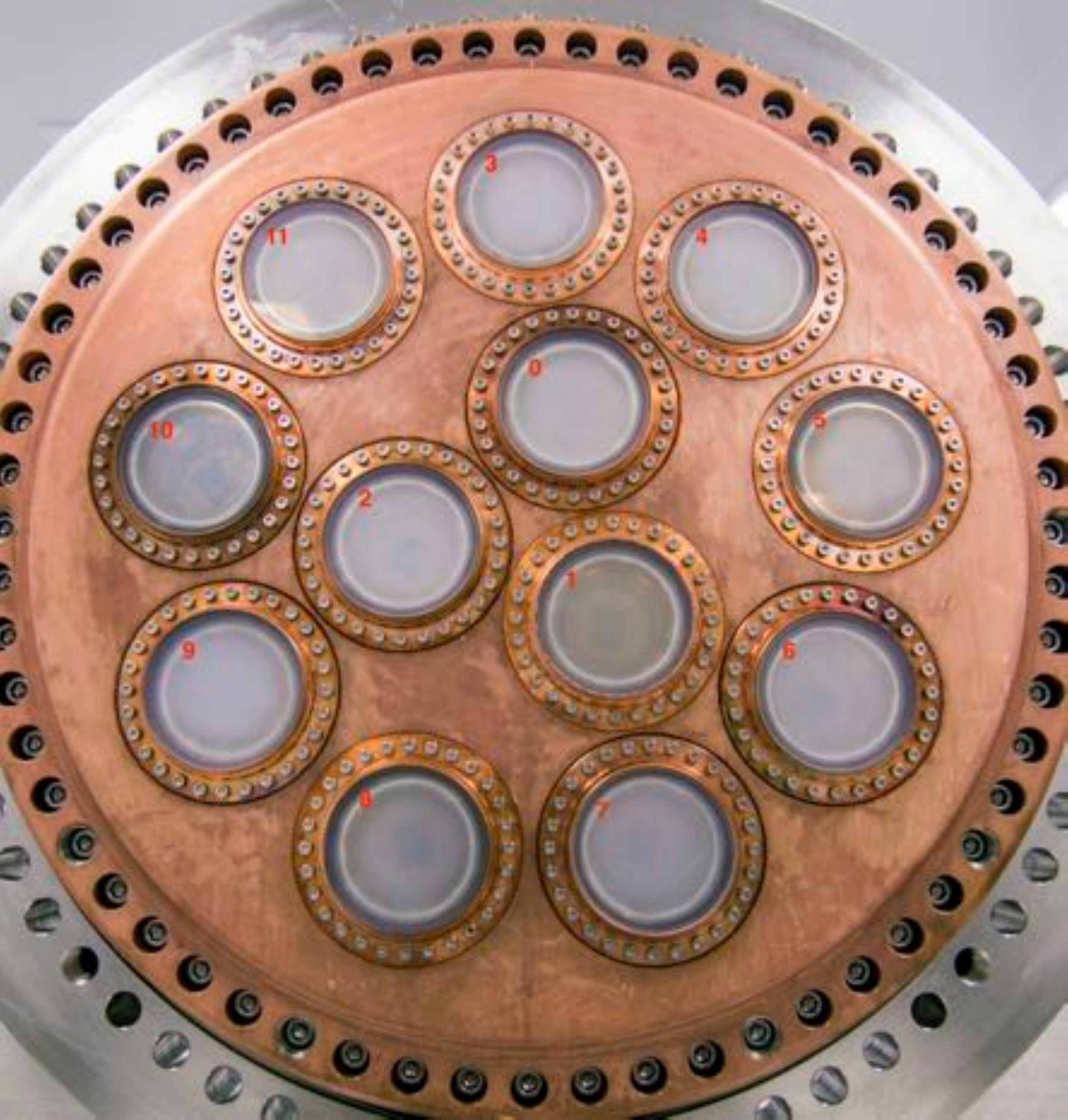}
  \end{center}
  \caption{Left: Relative charge seen by each PMT using X-rays of low energy emitted by radioactive sources; Right: a picture of the energy plane showing differences in coating quality.}
  \label{fig:PoisM2}
\end{figure}

The mean of the Poisson statistics relating the different Gaussians used in the fit gives an indication of the average number of detected photons. Monitoring this value for the same LED settings and the relative value for all runs yields an estimation of the relative PDE of the PMT systems (this includes the response of the PMT itself, plus the transmittance of the sapphire window coated with TPB and PEDOT) and its stability respectively. As can be seen in figure \ref{fig:PoisM}, the values vary notably between the different sensors, however those relative to that of the sensor with the highest value are consistent between runs. The relevance of this observation can be seen when one compares the relative values with the relative charge seen in each PMT for point-like events in the detector. Figure \ref{fig:PoisM2} shows that the relative charge observed follows the same pattern as the relative Poisson mean. As can be seen in the photograph taken at the time of installation (\ref{fig:PoisM2} right panel), the PMTs generally seeing less light are those with poorer quality deposition on the sapphire window suggesting absorption in the coating layers, or those in the outer ring. Since all calibrations were performed using the same LED, there is a possible contribution due to the position of the LED (the PMT which registers most light is the closest to the position in front of the LED) but this seems to be a small contribution since the X-ray data from radioactive sources are generally in good agreement with the calibrations.

\subsection{Calibration of the tracking plane}

\subsubsection{SiPM connectivity}

\begin{figure}[!htb]
  \begin{center}
    \includegraphics[width=0.49\textwidth]{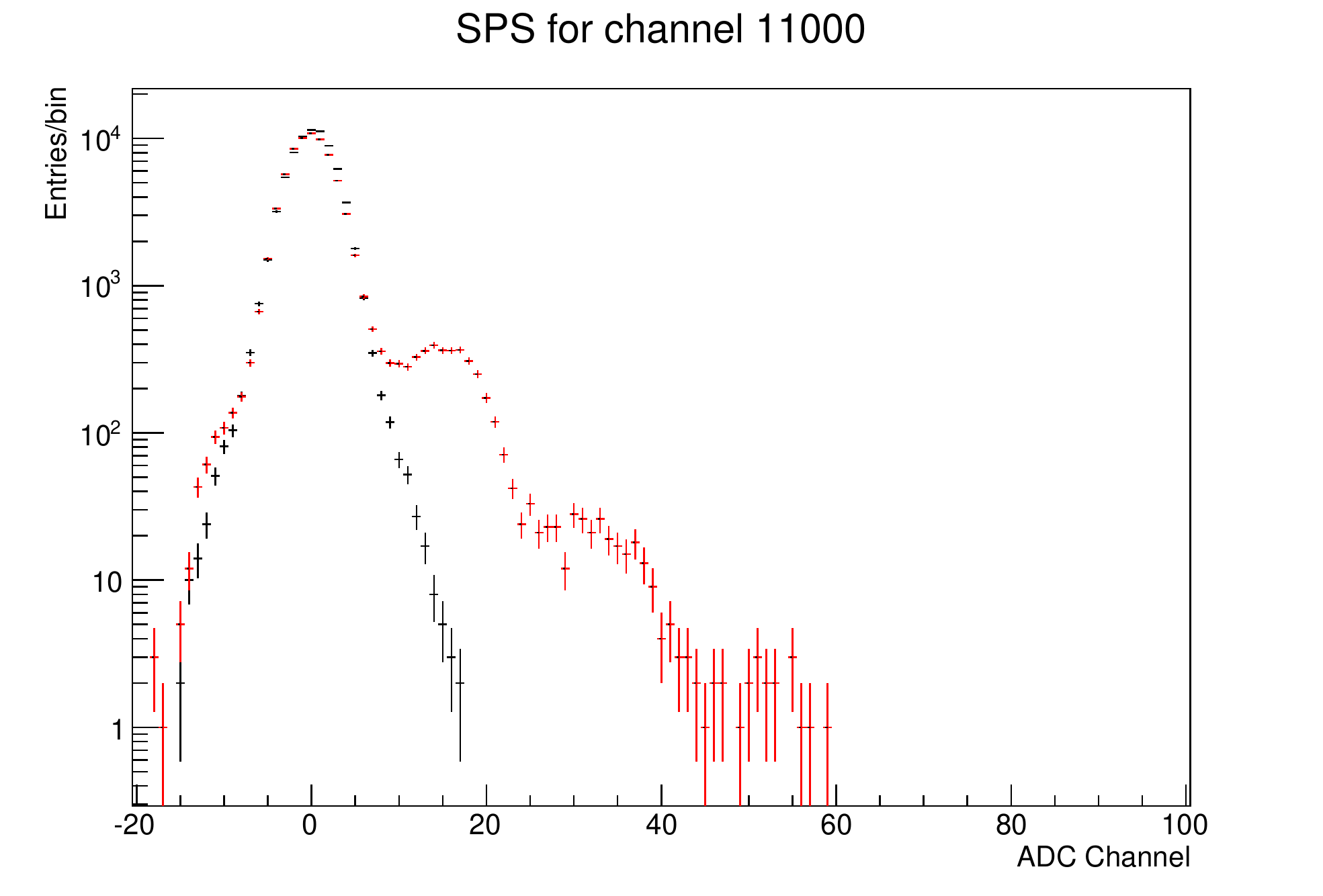}
    \includegraphics[width=0.49\textwidth]{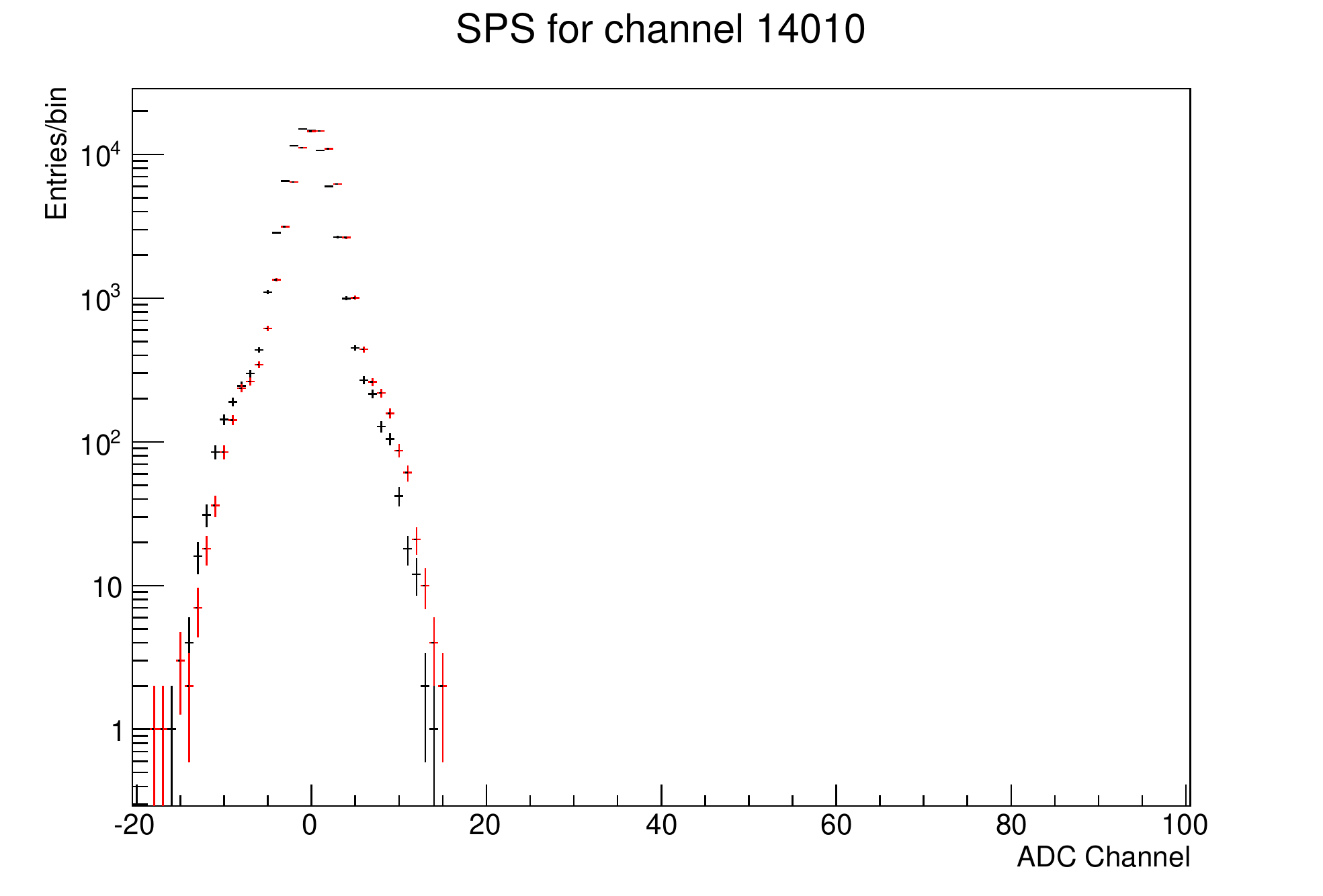}
  \end{center}
  \caption{Left: a normal channel showing a dark count spectrum in red
  and an electronics only spectrum in black; Right: a faulty channel.}
  \label{fig:connect}
\end{figure}

Of the \NewNumberOfSiPM\ channels of the tracking plane, only \NewDeadSiPM\ have been identified
as unresponsive with a further \NewUnstableSiPM\ showing some
instability. Additionally, there are \NewOutSiPM\ of the sensors just outside of
the light tube which do not see sufficient LED light for a reliable
calibration. As such, over \NewTrackingPlaneActiveAndStable\ of the tracking plane is active and
stable.

Connectivity is monitored at each calibration run by comparing the spectra
seen with and without bias voltage and identifying those with little
or no difference or with no obvious dark count peaks as faulty. Figure
\ref{fig:connect} shows examples of faulty and normal channels.

\subsubsection{SiPM Gain}

\begin{figure}[!htb]
  \begin{center}
    \includegraphics[width=0.49\textwidth]{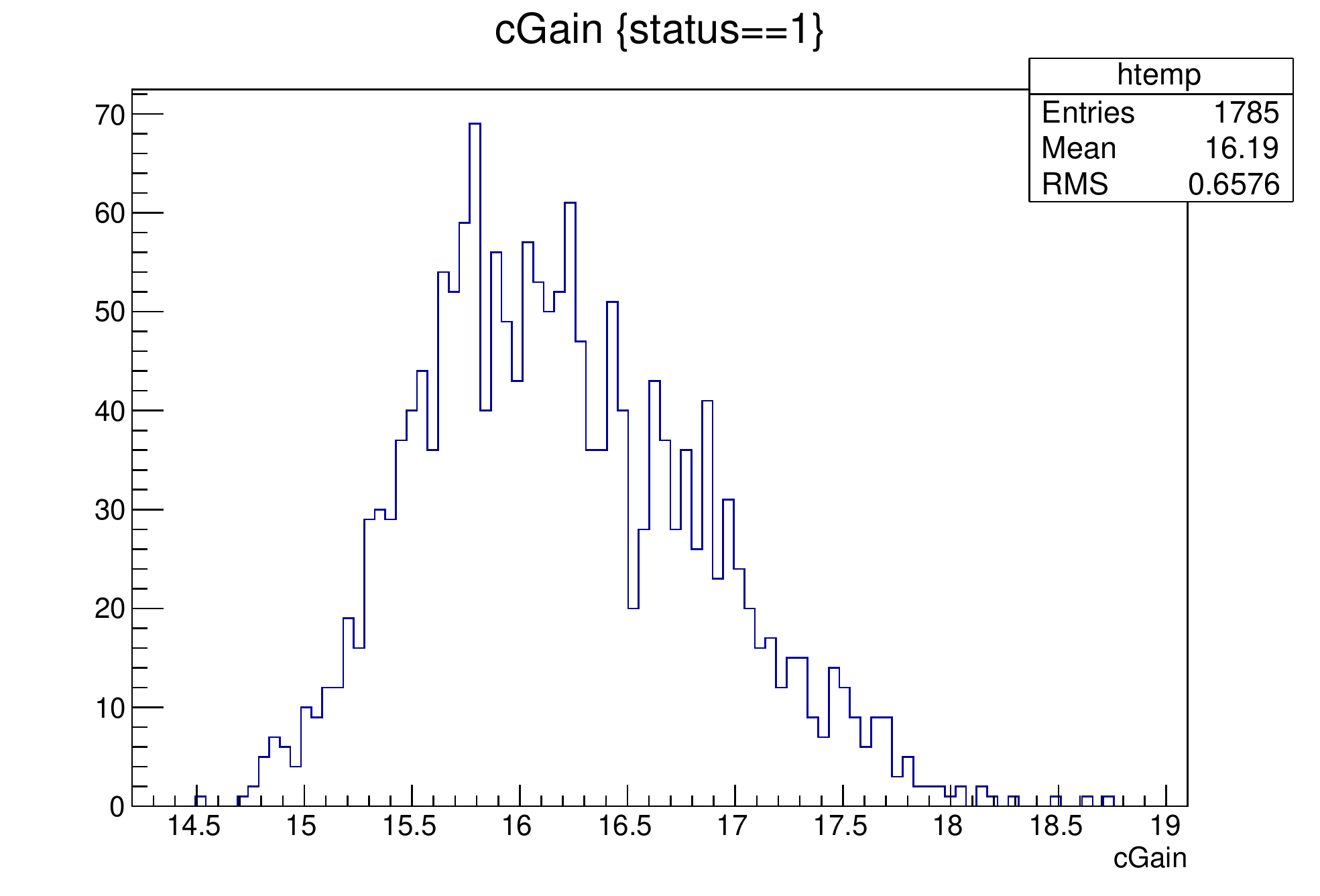}
    \includegraphics[width=0.49\textwidth]{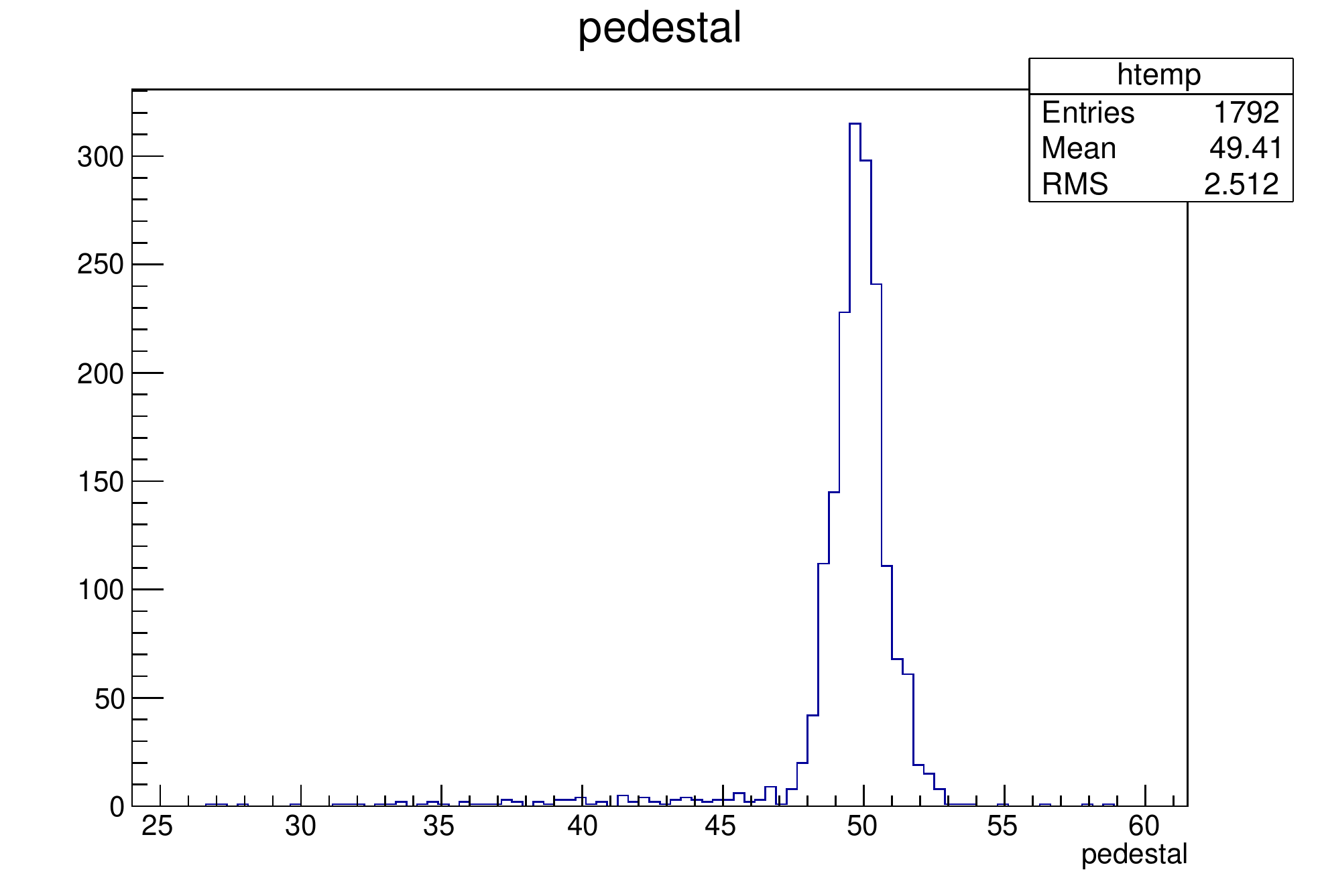}\\
    \includegraphics[width=0.49\textwidth]{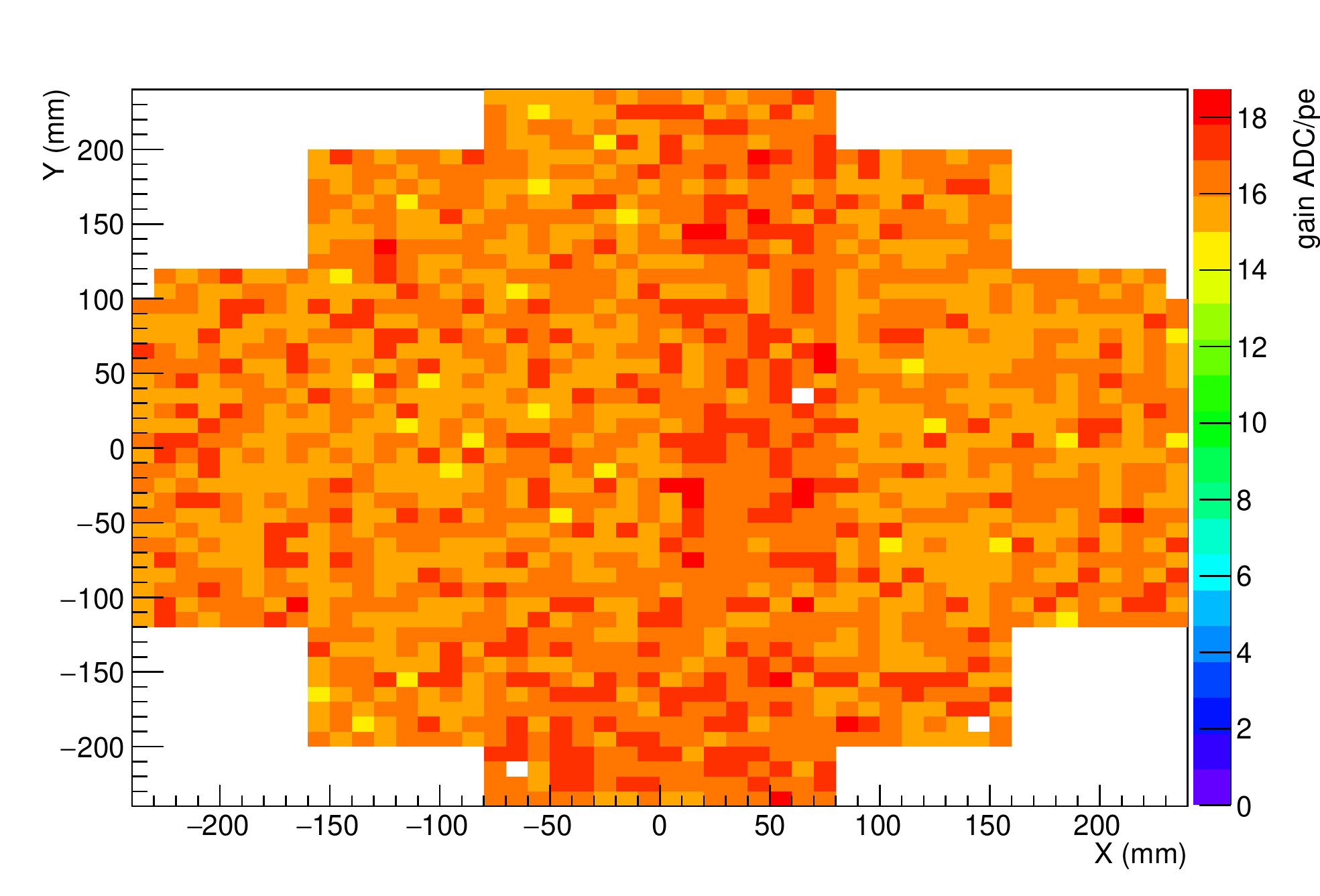}
  \end{center}
  \caption{SiPM calibration. Left: the gain distribution for the
  tracking plane sensors; Right: the average pedestals. Bottom:
  gain as a function of XY position.}
  \label{fig:Sigain}
\end{figure}

%

The single photon response of the sensors of the tracking plane is
determined using the light from a pulsed LED located at the energy plane.
The gain is then determined using the distance between the pedestal
peak and those for 1, 2, 3... spes. The data are processed by calculating
the baseline in each event and taking
\SI{2}{\mu s} integrals at the trigger time. Spectra are built from the baseline-subtracted
integrals and then fitted for the gain. The
calibration results in average pedestals for each channel as well as
the single photon conversion values. The results from a recent
calibration are shown in figure \ref{fig:Sigain}.
The calibration of the sensors has remained stable over the whole run
with only slight variations between runs.

%% file: src/detector_operation.tex
\section{Initial operation of the NEW detector: Run II}
\label{sec.operation}

\begin{figure}[!htb]
\centering
\includegraphics[angle=0, width=0.99\textwidth]{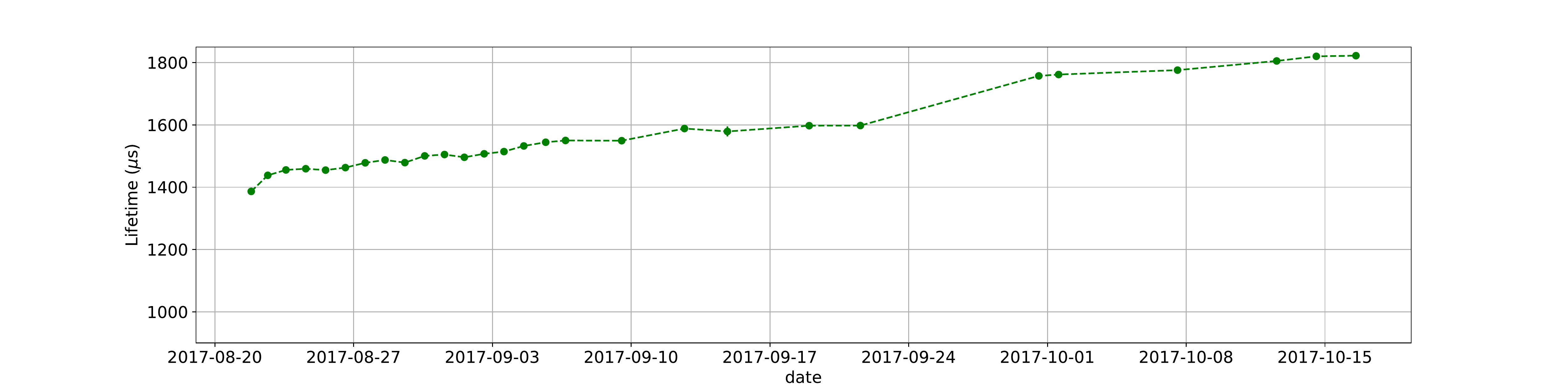}
\caption{Evolution of the lifetime during a fraction of Run II.} 
\label{fig.lifetime}
\end{figure}

The detector started operations at the LSC late in 2016. After a short engineering run (Run I) in November-December 2016, the detector was operated continuously between March and December, 2017 (Run II). 

In order to assess the overall performance of the detector and the gas system while keeping a safety margin (\eg\ minimizing the effect of possible leaks as well as limiting the energy of potential sparks in the gate), the operational pressure in Run II was limited to \NewSevenBarPressureRunII\ during the first part of the data taking. The drift voltage was set at \NewDriftField\  and the reduced field in the EL region at \NewReducedFieldRunII, slightly below the nominal value of \NewReducedField. Thus, the gate voltage was kept at \NewGateVoltageRunII\ and the cathode voltage at
\NewCathodeVoltageRunII. Under those conditions, the chamber was extremely stable, with essentially no sparks recorded over the period.

During the second part of the run the pressure was raised to  \NewNineBarPressureRunII, and the voltages were correspondingly adjusted to keep the same drift voltage and reduced field than during the first part of the run.  


Electron lifetime 
was low
at the beginning of \RII\, possibly due to a small oxygen leak. This led to adding an exhaustive helium leak check of the full system before filling the chamber with xenon, which has been implemented for Run III, currently underway.  On the other hand, the lifetime of the system could be improved by continuous gas circulation through the getters to an acceptable value of 2 ms. \Fig\ \ref{fig.lifetime} shows the evolution of the lifetime during a fraction of the run. It is expected that the combination of a stringent detector tightness (guaranteed through the now-standard helium test) and getter circulation will allow even larger lifetimes for Run III and beyond. 

Run II was mostly devoted to calibrations with radioactive sources, including \KR\, \NA, \CS\ and \TL\ sources. The primary goal was the overall characterization of the chamber, and an initial assessment of the energy resolution and topological signature. A first estimation of the background level in the detector was also made. 


%% file: pool/ack2018.tex
The NEXT Collaboration acknowledges support from the following agencies and institutions: the European Research Council (ERC) under the Advanced Grant 339787-NEXT; the Ministerio de Econom\'ia y Competitividad of Spain under grants FIS2014-53371-C04, the Severo Ochoa Program SEV-2014-0398 and the Mar\'ia de Maetzu Program MDM-2016-0692; the GVA of Spain under grants PROMETEO/2016/120 and SEJI/2017/011; the Portuguese FCT and FEDER through the program COMPETE, projects PTDC/FIS-NUC/2525/2014 and UID/FIS/04559/2013; the U.S.\ Department of Energy under contracts number DE-AC02-07CH11359 (Fermi National Accelerator Laboratory), DE-FG02-13ER42020 (Texas A\&M) and de-sc0017721 (University of Texas at Arlington); and the University of Texas at Arlington. We acknowledge partial support from the European Union Horizon 2020 research and innovation programme under the Marie Sklodowska-Curie grant agreements No. 690575 and 674896. We also warmly acknowledge the Laboratorio Nazionale di Gran Sasso (LNGS) and the Dark Side collaboration for their help with TPB coating of various parts of the \NEW\ TPC. Finally, we are grateful to the Laboratorio Subterr\'aneo de Canfranc for hosting and supporting the NEXT experiment.